\documentclass[fleqn,usenatbib]{mnras}

\usepackage{newtxtext} 
\usepackage{mathptmx}
\usepackage{txfonts}
\usepackage{float}

\usepackage[T1]{fontenc}
\usepackage{xcolor} 
\DeclareRobustCommand{\VAN}[3]{#2}
\let\VANthebibliography\thebibliography
\def\thebibliography{\DeclareRobustCommand{\VAN}[3]{##3}\VANthebibliography}

\usepackage{graphicx}	
\usepackage{amssymb}	
\usepackage{float}

\DeclareRobustCommand{\ion}[2]{%
\relax\ifmmode
\ifx\testbx\f@series
{\mathbf{#1\,\mathsc{#2}}}\else
{\mathrm{#1\,\mathsc{#2}}}\fi
\else\textup{#1\,{\mdseries\textsc{#2}}}%
\fi}

\title[Quiescent or dusty? Unveiling the nature of extremely red galaxies at $\mathrm{z>3}$]{Quiescent or dusty? Unveiling the nature of extremely red galaxies at $\mathrm{z>3}$}
\author[L. Barrufet et al.]{L. Barrufet$^{1,2}$\thanks{E-mail:lbarrufet@roe.ac.uk}, 
P. A. Oesch$^{2,3}$, 
R. Marques-Chaves$^{2}$, 
K Arellano-Cordova$^{1}$,
J.F.W. Baggen$^{4}$, 
A. C. Carnall$^{1}$, \newauthor
F. Cullen$^{1}$, 
J. S. Dunlop$^{1}$,
R. Gottumukkala$^{2,3}$, 
Y. Fudamoto$^{5}$
G. D. Illingworth$^{6}$, 
D. Magee$^{6}$, 
R. J. McLure$^{1}$, \newauthor
D. J. McLeod$^{1}$, 
M. J. Micha{\l}owski$^{7,1}$, 
M. Stefanon$^{8,9}$,  
P. G. van Dokkum$^{4}$, 
A. Weibel$^{2}$
\\ 
$^{1}$ Institute for Astronomy, University of Edinburgh, Royal Observatory, Edinburgh, EH9 3HJ \\
$^{2}$ Department of Astronomy, University of Geneva, Chemin Pegasi 51, 1290 Versoix, Switzerland \\
$^{3}$ Cosmic Dawn Center (DAWN), Niels Bohr Institute, University of Copenhagen, Jagtvej 128, K\o benhavn N, DK-2200, Denmark \\
$^{4}$ Department of Astronomy, Yale University, New Haven, CT 06511, USA \\
$^{5}$ Center for Frontier Science, Chiba University, 1-33 Yayoi-cho, Inage-ku, Chiba 263-8522, Japan \\
$^{6}$ Department of Astronomy and Astrophysics, University of California, Santa Cruz, CA 95064, USA \\
$^{7}$Astronomical Observatory Institute, Faculty of Physics, Adam Mickiewicz University, ul.~S{\l}oneczna 36, 60-286 Pozna{\'n}, Poland \\
$^{8}$ Departament d'Astronomia i Astrof\`isica, Universitat de Val\`encia, C. Dr. Moliner 50, E-46100 Burjassot, Val\`encia,  Spain \\
$^{9}$ Unidad Asociada CSIC "Grupo de Astrof\'isica Extragal\'actica y Cosmolog\'ia" (Instituto de F\'isica de Cantabria - Universitat de Val\`encia)
}

\date{Submitted April 2024}

\pubyear{2024}

\begin{document}
\label{firstpage}
\pagerange{\pageref{firstpage}--\pageref{lastpage}}
\maketitle

\begin{abstract} 
The advent of the {\it JWST} has revolutionised our understanding of high-redshift galaxies. In particular, the NIRCam instrument on-board {\it JWST} has revealed a population of {\it Hubble Space Telescope} ({\it HST})-dark galaxies that had previously evaded optical detection, potentially due to significant dust obscuration, quiescence, or simply extreme redshift. Here, we present the first NIRSpec spectra of 23 {\it HST}-dark galaxies ($\mathrm{H-F444W>1.75}$), unveiling their nature and physical properties. This sample includes both dusty and quiescent galaxies with spectroscopic data from NIRSpec/PRISM, providing accurate spectroscopic redshifts 
with $\mathrm{\overline{z}_{spec} = 4.1 \pm 0.7}$. The spectral features demonstrate that,  while the majority of {\it HST}-dark galaxies are dusty, a substantial fraction, $\mathrm{13^{+9}_{-6} \%}$, are quiescent. For the dusty galaxies, we have quantified the dust attenuation using the Balmer decrement ($\mathrm{H\alpha / H\beta}$), finding attenuations $\mathrm{A_{V} > 2\ mag}$. We find that {\it HST}-dark dusty galaxies are  $\mathrm{H\alpha}$ emitters with equivalent widths spanning the range  $\mathrm{  68 \AA <  EW_{H\alpha} < 550 \AA }$, indicative of a wide range of recent star-formation activity. 
Whether dusty or quiescent, we find that {\it HST}-dark galaxies are predominantly massive, with 85\% of the galaxies in the sample having masses $\mathrm{log(M_{*}/M_{\odot}) > 9.8}$. 
This pilot NIRSpec program reveals the diverse nature of {\it HST}-dark galaxies and highlights the effectiveness of NIRSpec/PRISM spectroscopic follow-up in distinguishing between dusty and quiescent galaxies and properly quantifying their physical properties. Upcoming research utilising higher-resolution NIRSpec data and combining {\it JWST} with ALMA observations will enhance our understanding of these enigmatic and challenging sources.

\end{abstract}

\begin{keywords}
Galaxies: high-redshift. Infrared: galaxies. Techniques: spectroscopic

\end{keywords}

\section{Introduction}
\label{Introduction}

Over the past decade, infrared observations using the {\it Spitzer Space Telescope} \citep[e.g.,][]{Caputi2012,Wang2016,AlcaldePampliega2019}
and mm-wavelength detections in Atacama Large Millimeter/submillimeter Array (ALMA) continuum data \citep[e.g.,][]{Franco2018,Yamaguchi2019,Williams2019,Wang2019,Gruppioni2020,Xiao2023a}
have hinted at a significant population of dust-obscured galaxies that were missing from {\it Hubble Space Telescope} ({\it HST}) datasets. These objects have become known as `{\it HST}-dark' galaxies. However, due to the constraints imposed by low spatial resolution observations (in the case of {\it Spitzer})
and the absence of spectroscopic redshifts, the reliable determination of their physical properties, including stellar masses and star-formation rates (SFRs), have remained extremely difficult. More recent studies with the {\it James Webb Space Telescope} ({\it JWST}) have now confirmed that massive, dusty, high-redshift galaxies were indeed missed from earlier (primarily optically-selected) samples \citep[e.g.,][]{Nelson2022,Barrufet2023,PerezGonzalez2023,Rodighiero2023,Labbe2023a,GomezGuijarro2023}.

The high-resolution multi-band $\lambda=1-5\,\mu$m photometry provided by {\it JWST}/NIRCam is transforming our understanding of these red galaxies. Expressly, it has now been confirmed that these sources are generally located at redshifts $\mathrm{z>3}$ and typically lie on the main sequence of star-forming galaxies \citep[e.g.][]{Barrufet2023, Rodighiero2023}. Substantial further progress is now also possible by using the spectroscopic capabilities of {\it JWST}/NIRSpec to probe the physical properties of {\it HST}-dark galaxies in more detail \citep{Williams2023, PerezGonzalez2024}.

A further discussion triggered by early {\it JWST} data concerns the potential for the existence of massive galaxies at high redshift to present a challenge to the standard $\Lambda$CDM cosmological model (e.g., \citealt{BoylanKolchin2023, Lovell2023}), although ongoing spectroscopic follow-up of several of the sources in question is already suggesting that such claims may have been premature. A notable case is that of CEERS-3210, a candidate early massive galaxy, initially reported as being at $\mathrm{z_{phot} \simeq 8}$ by \cite{Labbe2023a}, which has now been revealed by NIRSpec spectroscopy to in fact be an active galactic nucleus (AGN) at $\mathrm{z_{spec} = 5.624}$ \citep{Kocevski2023}. This example underscores the importance of securing spectroscopic redshifts for the accurate characterization of red galaxies in particular, given that the optical-infrared spectral energy distribution (SED) displayed by these sources often results in poorly constrained photometric redshifts, even with the best available NIRCam data. 

The precise characterisation of the number and nature of distant red galaxies is also of critical importance for the accurate determination of the high-mass end of the high-redshift galaxy stellar mass function (\citealt{Gottumukkala2024}, see also \citealt{Weibel2024}). While the current revolution in our understanding of early galaxy evolution is undoubtedly being driven primarily by {\it JWST}/NIRCam, it is already clear that imaging with the {\it JWST} Mid-Infrared Instrument (MIRI) camera at $\lambda>5\,\mu$m is also invaluable. In particular, MIRI has been pivotal for confirming the presence of massive galaxies with significant dust attenuation at $\mathrm{z \simeq 7-8}$ \citep{Akins2023, Barro2023}. Moreover, it has also been shown that the inclusion of MIRI data can significantly improve the constraints on inferred stellar masses, in some cases yielding more moderate values \citep{Williams2023}, which suggests that early galaxies do not present a serious challenge to our current understanding of cosmology \citep{Wang2024}. 
In parallel, \citet{PerezGonzalez2024} have underscored the importance of MIRI data for determining whether the continuum emission from red sources, including the compact objects now frequently referred to as `Little Red Dots' (LRDs) \citet{Matthee2023b}, is dominated by emission from an obscured AGN or starlight. 

In addition to confirming the existence of dusty galaxies at high redshift, {\it JWST} is also revealing significant numbers of massive galaxies in the early Universe which are red because their stellar populations are already significantly evolved at $z \simeq 3-5$, implying that star-formation activity in these galaxies was somehow terminated at very early times \citep[e.g.,][]{Carnall2023,Carnall2023b,Valentino2023}. 
These galaxies are red primarily because they exhibit prominent breaks in their spectra at rest-frame $\lambda_{\rm rest}\simeq4000\AA$. However, it can still be very challenging to distinguish them from dusty galaxies when deep imaging is only available at ${\rm \lesssim2\,\mu m}$. While these objects often appear bluer at longer wavelengths, meaning that in principle, they can be distinguished from their dusty counterparts using NIRCam+MIRI photometry, recent findings also highlight the existence of galaxies at $z\simeq4$ that are both quiescent and dusty (\citealt{Setton2024}). The accurate identification of quiescent galaxies at $\mathrm{z>3}$ is crucial for precisely determining their number density, and hence advancing our understanding of the onset and subsequent quenching of star-formation activity at very early times  \citep[i.e.,][]{Merlin2019,Santini2021}. 

The advent of extensive {\it JWST} imaging surveys, as exemplified by the Public Release Imaging for Extragalactic Research (PRIMER) survey (PI: J. Dunlop), heralds a new era of statistically meaningful studies of red galaxies (both quiescent and dusty) due to the power of extensive NIRCam+MIRI imaging over the best-studied {\it HST} legacy fields (Dunlop et al. in prep., Barrufet et al. in prep.).
However, even with expanded NIRCam+MIRI imaging data, it is clear that spectroscopic follow-up with {\it JWST}/NIRSpec will be crucial for the study of red galaxies in particular. Such observations are essential for delivering reliable and precise spectroscopic redshifts for such objects, breaking the photometric redshift degeneracies often arising from their red SEDs. Moreover, the broad wavelength coverage and moderate spectral resolution provided by the NIRSpec PRISM mode, in particular, provides both a full suite of emission lines (where present), as well as high signal-to-noise ratio (SNR) detections of continuum emission (and hence, potentially,  absorption lines), enabling dusty and quiescent red galaxies to be unambiguously distinguished. NIRSpec follow-up spectroscopy is thus urgently required to resolve the outstanding questions posed by the latest advances in the study of red galaxies, which have arisen from the deep {\it JWST} imaging observations. 

In this rapidly evolving context, we here report the first results from the Cycle-1 NIRSpec pilot programme, `Quiescent or Dusty? Unveiling the Nature of Red Galaxies at $\mathrm{z>3}$' (Programme ID: 2198; PIs: L. Barrufet \& P. Oesch), providing the first {\it JWST} spectroscopic follow-up data focused specifically on {\it HST}-dark galaxies. For the reasons explained above, the key aims of this program are to i) deliver spectroscopic redshifts, ii) enable the classification of individual HST-dark galaxies as either quiescent or dusty, and iii) provide robust estimates of their basic physical properties, including dust attenuation and stellar masses.

The paper is structured as follows. In Section \ref{observations}, we describe the design of our NIRSpec programme and provide details of the source catalogue produced from our NIRCam pre-imaging (through the F444W and F200W bands). Our multi-band photometric catalogue is completed with photometry from First Reionization Epoch Spectroscopically Complete Observations \citep[FRESCO;][]{Oesch2023}, and from the {\it JWST} Advanced Deep Extragalactic Survey \citep[JADES;][]{Eisenstein2023}.  
The methodology we adopted to analyse our spectroscopic data is presented in Section \ref{methodology}, and then our spectroscopic results 
are presented in Section \ref{Spectra_section}, where a distinction between quiescent and dusty categories is made, and the results of this classification are discussed in the context of observed galaxy morphology. The derived physical properties of the {\it HST}-dark galaxies are presented in Section \ref{physical_properties}, and we then discuss our results in Section \ref{Discussion}. Finally, we summarise our conclusions in Section \ref{Summary}.

Throughout, all magnitudes are given in the AB system \citep[e.g.,][]{Oke83}, and for all cosmological calculations we adopt $\mathrm{\Omega_{M} =0.3}$, $\mathrm{\Omega_{\Lambda} =0.7}$ and $\mathrm{H_{0}= 70\,km \ s^{-1} \ Mpc^{-1}}$.

\section{Observations}
\label{observations}

As outlined above, the primary aim of our programme `Quiescent or dusty? Unveiling the nature of red galaxies at $\mathrm{z>3}$'  (GO-2198; PIs: L. Barrufet \& P. Oesch) was to obtain low-resolution  NIRSpec PRISM spectra of a representative sample of {\it HST}-dark galaxies to determine the redshifts and nature of these enigmatic galaxies. In this section, we describe how our targets were selected and summarize the basic reduction and analysis of the NIRSpec data, which cover an observed wavelength range 1-5 $\mathrm{\mu m}$ at a typical spectral resolution of $\mathrm{R=100}$.

As we describe further below, we were able to target 23 galaxies which qualify as {\it HST}-dark, which are the focus of this paper. However, our two NIRSpec pointings meant that we were able to obtain spectra of $\mathrm{\sim 140}$ objects with the NIRSpec Micro-Shutter Assembly (MSA). To fully exploit this observational opportunity, we therefore selected $\mathrm{\sim 110}$ `fillers', which we selected as potential massive galaxies at $\mathrm{z>3}$ from the 3D-HST catalogue \citep{Skelton2014}. 

\subsection{(Pre-)Imaging Data and Photometric Catalogue}

Our NIRSpec program was designed to follow up on several HST-dark galaxies that were initially identified in Spitzer/IRAC imaging and remained undetected in HST imaging data in the GOODS-S field from the GREATS survey \citep{Labbe2015,Stefanon2021}. The program's main targets could be covered with two pointings of NIRSpec. 

Since the IRAC point-spread function is much too wide to pinpoint the accurate location of these galaxies, JWST NIRCam pre-imaging was required. This pre-imaging was taken between November 10-19, 2022, in the F444W and F200W filters. The short exposures were taken with the BRIGHT2 readout pattern with seven groups/integration and a 6-point FULLBOX dither pattern, leading to a total exposure time of 901s. For each of the two pointings, the images were mosaicked in two rows to span the entire field-of-view of NIRSpec. 

The images were processed and aligned with the \texttt{grizli} tool, before deriving multi-band photometric catalogs. In addition to the JWST/NIRCam images, pre-existing HST imaging with ACS and WFC3/IR was included. Finally, a small portion of each field overlapped with FRESCO, where public F182M, F210M, and deeper F444W imaging were available for the catalogues used for creating the masks back in November 2022. Since then, a good part of the field was additionally covered by deep JADES imaging both from the original GTO program as well as from the cycle 2 JADES Origins Field \citep{Eisenstein2023,Eisenstein2023b}. Hence, several of our targets are now covered by deep multi-band photometry, including medium-band filters. 

For the photometric catalogues adopted in the following, we use all the v7 ACS, WFC3/IR, and NIRCam images available in the GDS-SW field on the public DAWN JWST Archive (DJA)\footnote{\url{https://dawn-cph.github.io/dja/imaging/v7/}}. Since these do not yet include the pre-imaging from this program, this was added in a separate stack of the F200W and F444W data. This ensures that all sources are covered by at least two NIRCam bands. All images were aligned to the same 40mas pixel grid.  

To construct photometric catalogues, we follow the same procedure as outlined in
\citealt[][see also, e.g., \citealt{Barrufet2023,Gottumukkala2024}]{Weibel2024}. 
Briefly, \texttt{SExtractor} \citep[][]{Bertin96} is run in dual image mode using a PSF-matched, inverse-variance weighted stack of the F277W, F356W and F444W images as the detection image. The remaining filters were PSF-matched to F444W using PSFs created directly from stars in the images. Photometry was measured in small circular apertures of 0\farcs48 diameter and corrected to total fluxes using Kron AUTO apertures from the F444W images plus a small residual correction for flux lost from the wings of the PSF.

As a general quality cut, the catalogue only contains objects with S/N $> $5 in at least one of the available NIRCam-wide filters. We also removed objects near the edge of the image, stars, and spurious detections. Finally, to account for possible systematic uncertainties in the photometry, we apply an error floor of 5 \% to all flux measurements before running any SED fitting.

\subsection{Spectroscopic data}
\label{spectroscopi-catalogue}

This program contains two NIRSpec pointings that allowed us to target 140 galaxies with PRISM spectra at $R\sim100$. 
In each pointing, two MSA slit configurations were used for observations. These were observed on January 19, 2023 and February 6, 2023, respectively. The second pointing suffered NIRSpec shutter glitches affecting the spectra of the main sample which has been removed in this study. For each MSA mask, the spectra were taken with the PRISM grating and an NRSIRS2RAPID readout pattern of 55 groups/integration. We used three slitlets per target and dithered along the three shutters. Each spectrum thus received a total exposure time of 2451 s. This allows us to reach the continuum for faint sources and probes emission lines down to $\sim1\times\mathrm{10^{-18}\,erg\,s^{-1}\,cm^{-2}}$ (5$\sigma$).

The 140 NIRSpec spectra were reduced and extracted with \texttt{msaexp}\footnote{\url{https://github.com/gbrammer/msaexp/}}, a tool for extracting JWST NIRSpec MSA spectra. Individual slits were identified, considering the three shutter positions. After subtracting the sky background, the data were collapsed into a single 2D spectrum. This was then processed to extract the 1D spectra. Our data is identical to that available online on the DJA\footnote{\url{https://dawn-cph.github.io/dja/}}, except for source ID=7151, which was processed manually for improved slit extraction. 

Spectroscopic redshifts were also determined using the MSAExp tool. The wide wavelength range covered by the NIRSpec/PRISM spectra ($\mathrm{ \sim 1 - 5 \mu m}$) allowed us to determine an accurate spectroscopic redshift for every source in our mask, as they encompassed several spectral lines or the continuum with a Balmer Break for all sources. 

The 2D and 1D spectra of the main sample of this study, including their IDs and the spectroscopic redshifts of the HST dark galaxies, are presented in Appendix \ref{Appendix}. 

\begin{figure}
\centering
     \includegraphics[width=\columnwidth]{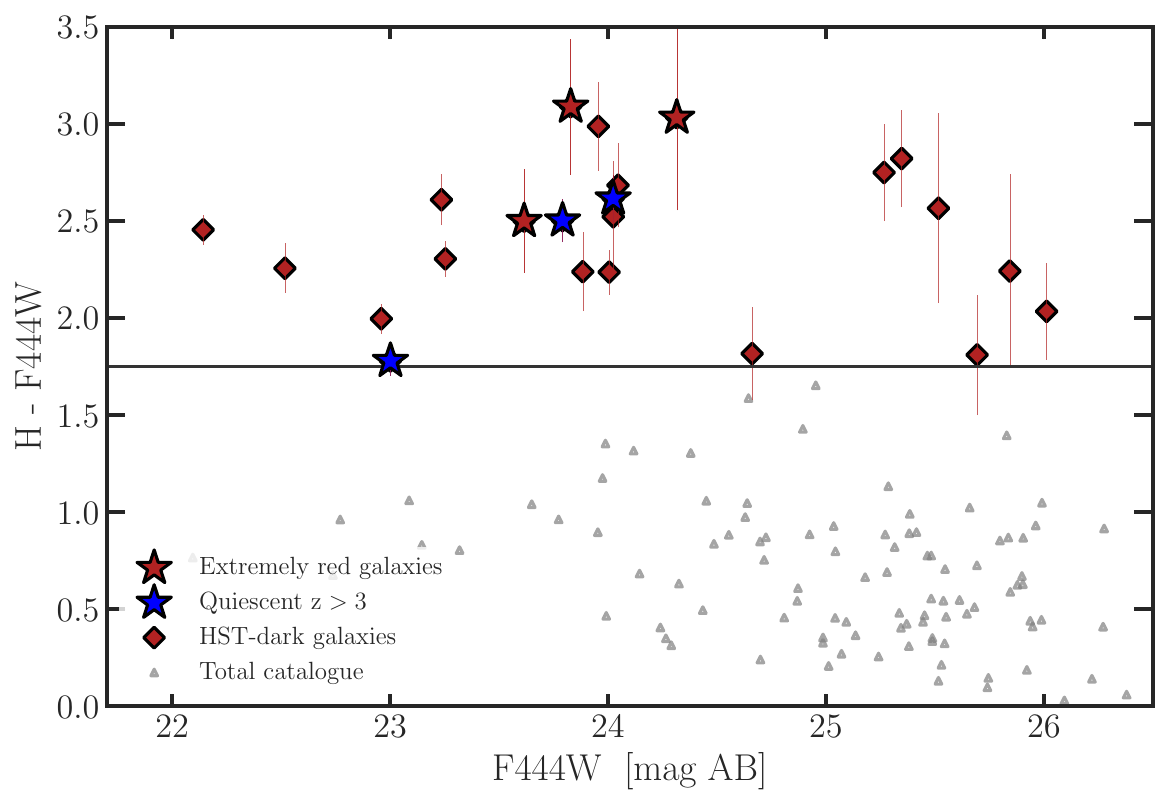} 
     \caption{Colour magnitude diagram (F444W vs H-F444W) of the total sample (grey triangles) of the  'Quiescent or dusty?' program (GO 2198). We selected 23 red galaxies (red diamonds) with the colour $\mathrm{H-F444W>1.75}$ similar to \citet{Barrufet2023} (black line). This subset includes several types of red galaxies: red stars represent those with extremely red spectra, while blue stars denote the three quiescent galaxies at $\mathrm{z>3}$, respectively indicating either significant dust presence or, conversely, a pronounced Balmer break signalling an old stellar population. }  
   \label{fig:colourselection}
\end{figure}

\subsection{Selection of HST-Dark galaxies}
\label{sec:colourselection}

The primary targets of this program are HST-dark galaxies. However, the masks were filled with other galaxies based on previous catalogues from 3D-HST \citep{Skelton2014} and our pre-imaging data. The total NIRSpec sample of 140 sources was classified by red colour using the HST H-band and the reddest band of NIRCam, F444W, at $\mathrm{1.6 \mu m}$ and $\mathrm{4.4 \mu m}$ respectively. Using a similar colour cut as employed in \citet{Barrufet2023} of $\mathrm{H-F444W > 1.75}$, a total of 23 galaxies in our MSA masks qualify as `HST-dark'. These are highlighted in Figure \ref{fig:colourselection}. Our primary targets span a broad range in F444W magnitude, between 22 and 26 mag. While these sources would be bright enough to be identified with Spitzer/IRAC data in these fields \citep[e.g.,][]{Stefanon2021}, only three among these sources were present in the previous catalogues of H-dropout galaxies from \citet{Wang2019}. In large part, this can be attributed to the $\sim10\times$ lower spatial resolution of the IRAC instrument compared to JWST/NIRCam data, but also to a redder colour cut of H-[4.5]$\gtrsim2$ in Spitzer-based HST-dark samples \citep[e.g.,][]{Wang2016}. 

With the NIRSpec spectra in hand, we can finally study the nature of these galaxies, which we do in the following sections. In Figure \ref{fig:colourselection}, we already highlight three distinct types of galaxies that are identified among the HST-dark sample that will be introduced and discussed in more detail in Section \ref{Spectra_section}.

\begin{figure}
\centering
\includegraphics[width=\columnwidth]{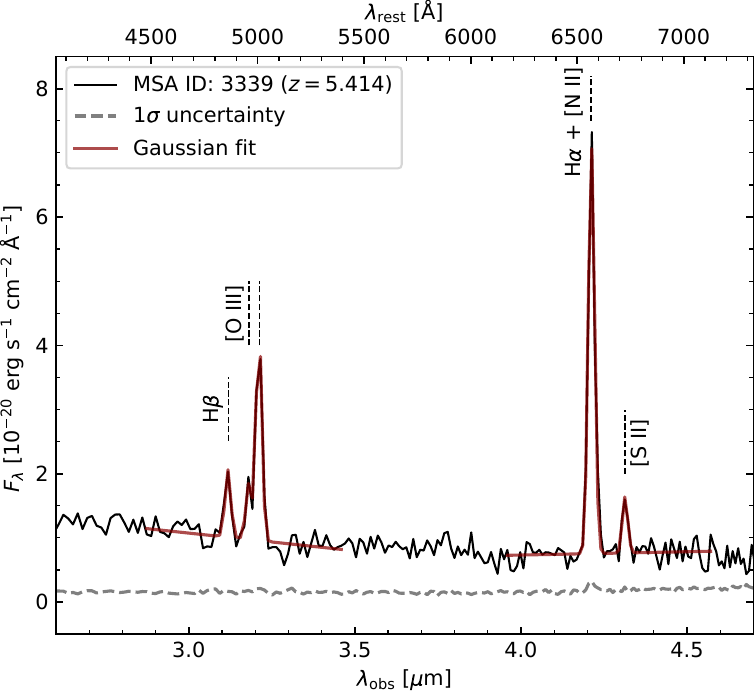} 
\includegraphics[width=\columnwidth]{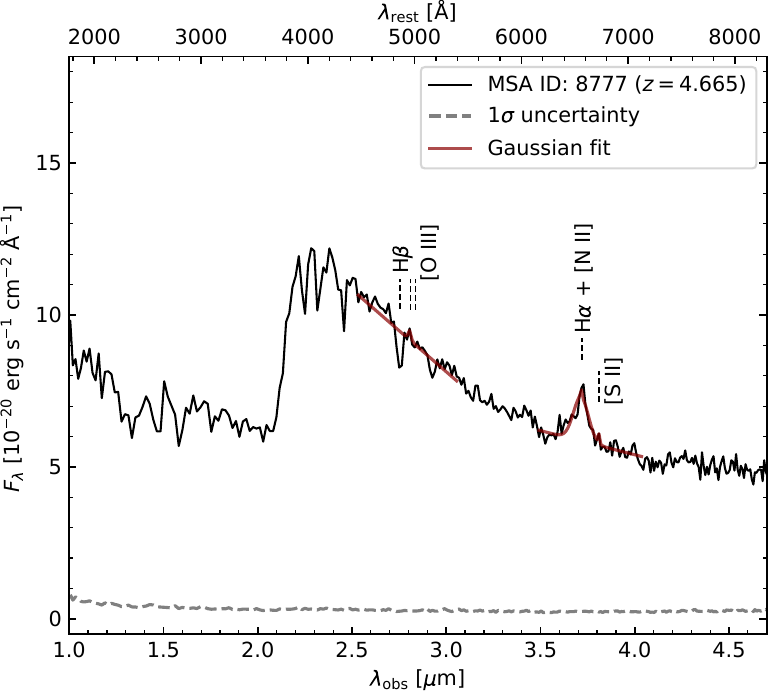} 
     \vspace{-3mm}
 \caption{This figure presents spectral fitting examples for two distinct HST-dark galaxies: dusty and quiescent. Black lines denote the observed spectra, while the red lines depict Gaussian fits to spectral lines and the linear fitting to the continuum. (Top) The dusty HST-dark galaxy (ID=3339) presents emission lines ($\mathrm{H \beta}$, $\mathrm{[OIII]}$, $\mathrm{H \alpha + [NII]}$ and $\mathrm{[SII]}$). (Bottom) The quiescent HST-dark galaxy (ID=8777) is characterised by its typical spectral profile and absence of emission lines, except for an unusual $\mathrm{H \alpha + [NII]}$ emission because of the AGN presence.  
 }  
   \label{3339_fittedspectra}
\end{figure}

\section{Methodology}
\label{methodology}

This section outlines the methodology employed to determine the physical properties of HST-dark galaxies, utilising spectroscopic data for classification and spectroscopic redshifts, alongside photometric data for quantifying stellar masses, SFRs and dust attenuation.

\subsection{Emission line extractions} 

We employed a direct visual examination of their spectra to classify HST-dark galaxies. HST-dark galaxies were identified as `quiescent' if they lacked emission lines and exhibited a pronounced Balmer Break. Conversely, galaxies were classified as `dusty' if their spectra showed dust-attenuation $\mathrm{A_{V} \gtrsim 1}$ mag. The red spectral slope is another evident sign of dustiness, indicating significant dust attenuation. 

The dusty HST-dark galaxies in our sample feature prominent $\mathrm{H\alpha}$+[NII] emission, with additional spectral lines like [SII], $\mathrm{H\beta}$, and [OIII] varying among galaxies. Hence, dust attenuation was assessed using the Balmer decrement between $\mathrm{H\alpha}$ and $\mathrm{H\beta}$, with spectroscopic redshifts obtained from PRISM spectral fitting (see Section \ref{spectroscopi-catalogue}). Utilizing these redshifts, we performed Gaussian fits to the $\mathrm{H\alpha}$ and $\mathrm{H\beta}$ lines to calculate their fluxes for dust attenuation assessment. More specifically, we simultaneously fit the spectral regions encompassing $\mathrm{H\beta+[OIII]}$ and 
$\mathrm{H\alpha+ [NII] + [SII]}$ using  Gaussian profiles for each line. We use a 400\AA\ spectral region for the continuum on both sides of these lines. Figure \ref{3339_fittedspectra} shows an example of the best-fit results for two of our targets, one dusty and a quiescent galaxy. 
However, when $\mathrm{H\beta}$ lines remained undetected, we used a conservative 2$\sigma$ upper limit. The continuum adjacent to these lines was fitted, and the line flux was subtracted.  Following this subtraction, we applied the method described by \citet{Dominguez2013} to determine the colour excess, using $\mathrm{E(B-V) = 1.97 \log((H\alpha/H\beta)/2.86)}$ based on a \citet{Calzetti2000} attenuation  curve. We converted from colour excess to dust attenuation $\mathrm{A_{v}}$ with the conversion factor applicable for starburst $\mathrm{R_{v} = 4.05}$. 

Note as a caveat that, due to the PRISM's resolution, $\mathrm{H\alpha}$ and [\ion{N}{ii}] $\lambda$6584 fluxes are blended, leading to an overestimation of $\mathrm{F_{H\alpha}}$. To minimize this effect,  we have inspected the contribution of [\ion{N}{ii}] to $\mathrm{F_{H\alpha}}$ by varying $\mathrm{10\%}$ of its contribution \citep[e.g.,][]{Sanders2015, Marmol-Queralto2016}. We find that the values of $A_{\rm v}$ change up to 0.37 dex, with non-critical impact in the resulting observed fluxes of H$\alpha$/H$\beta$. However, follow-up high-resolution observations of HST-dark galaxies are important to corroborate the contribution [\ion{N}{ii}] to the flux measurement of H$\alpha$.

Additionally, for $\mathrm{H\alpha}$ -- the sole line present in all dusty HST-dark galaxies -- we measured the line's full width at half maximum (FWHM) and the rest-frame equivalent width (EW). Given the low resolution of the prism spectra, however, 
the FWHM are not conclusive of broad lines except for two sources (see the results in Section \ref{Spectra_section}). 

\subsection{Spectral energy distribution fitting} 

We computed the physical properties of the 140 sources utilizing the multi-band photometric catalogue described in Section \ref{observations} and incorporating the spectroscopic redshifts described in Section \ref{spectroscopi-catalogue}. Specifically, we determined their star formation rates (SFRs), levels of dust attenuation, and stellar masses, emphasising the latter to evaluate if these sources are among the most massive observed in our sample. We apply the same methodology to both HST-dark galaxies and the remaining sample, facilitating a robust comparison. 

For this analysis, we employed the Bayesian Analysis of Galaxies for Physical Inference and Parameter EStimation (BAGPIPES) tool \citep{Carnall2018}, which constructs complex model galaxy spectra and fits them to a combination of spectroscopic and photometric data using the MultiNest nested sampling algorithm \citep{Feroz2019}. Importantly, unlike prior
NIRCam studies, our approach benefits from incorporating spectroscopic redshifts as an input, thereby reducing the degeneracies often associated with dual peaks in photometric redshift distributions caused by dust attenuation effects on redshift determination.

We have adopted similar model assumptions to those in \cite{Barrufet2023} and \cite{Gottumukkala2024}, which have effectively characterised HST-dark galaxies. 
We employ a delayed star formation history (SFH) model with an e-folding time ranging from $\mathrm{\tau = 0.1-9\ Gyr}$, using a uniform prior. This allows for the inclusion of both relatively short bursts and, effectively, a constant SFH. The stellar population models are based on the 2016 updated version of the \citet{Bruzual2003} library, using a \cite{Kroupa2001} initial mass function (IMF).
We consider a range of metallicities from 0.2 to 2.5 $Z_\odot$, taking the Solar value as $Z_\odot=0.02$. Nebular continuum and emission lines are included self-consistently, utilizing the photoionization code CLOUDY \citep{Ferland2017}, with the ionization parameter set to $\log U =-2$. Lastly, we apply the \citet{Calzetti2000} dust attenuation law, accommodating highly attenuated SEDs by setting an $\mathrm{A_{v}}$ range of ${\rm 0-6\,\ mag}$, using a uniform prior.

\begin{figure}
 \centering
     \includegraphics[width=\columnwidth]{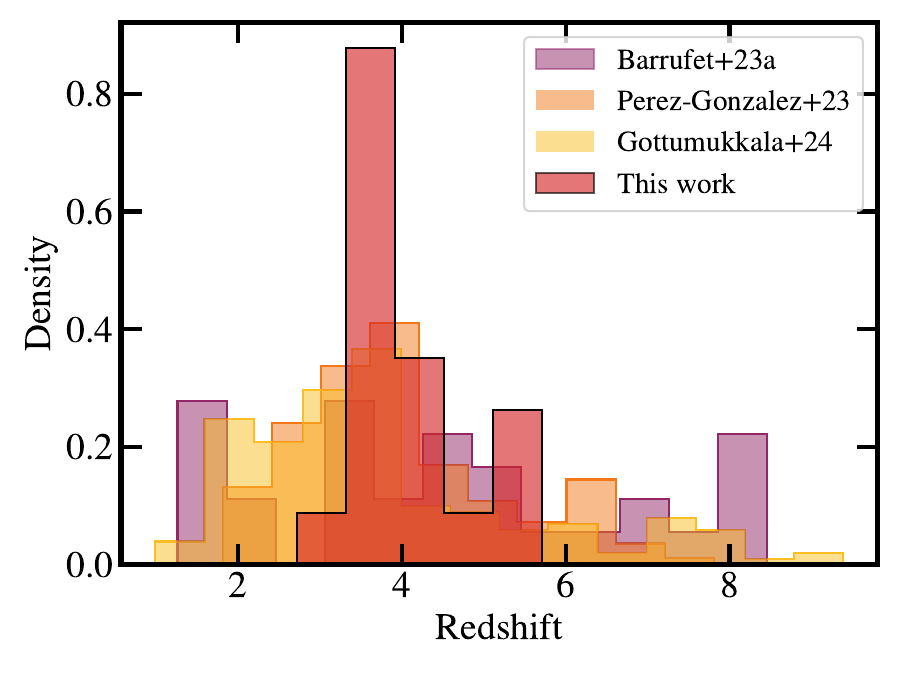}    
     \vspace{-3mm}
 \caption{Redshift distribution of HST-dark galaxies: this histogram presents our spectroscopically derived redshifts (red distribution) alongside photometric redshifts from comparative studies. Our data range from $\mathrm{z\sim2.7-5.5}$ and peak at $\mathrm{z \sim 4}$, underscoring our sample's mean redshift of HST-dark galaxies. Photometric data extend to higher redshifts beyond the scope of our pilot study, indicating the necessity for more in-depth observations.
 }  
   \label{HSTdark-histogram}
\end{figure}

\section{HST-dark galaxies Spectroscopic Results: Quiescent, Dusty, and AGN?}
\label{Spectra_section}

This section presents a first spectroscopic analysis of HST-dark galaxies spanning the $\mathrm{ \sim 1-5 \mu m}$ wavelength range, facilitated by JWST/NIRSpec observations. Additionally, we quantify the percentage of quiescent galaxies within our sample. 

Four of 23 analyzed HST-dark galaxies were excluded due to low-quality spectra, leaving 19 with high-quality spectra suitable for spectroscopic analysis. Despite these four spectra being compromised, noisy emission lines and photometric data still confirmed the galaxies' dusty nature. 

Even with NIRCam photometry, the photometric redshifts of HST-dark galaxies are difficult to determine due to the red SEDs of these galaxies. Therefore, the first question to address is simply the redshift distribution of these galaxies. This is shown in Figure \ref{HSTdark-histogram}, juxtaposed against earlier JWST/NIRCam photometric analyses.  We find a median redshift of $\mathrm{ \bar{z} = 4.1 \pm 0.7 }$, with all sources located at $\mathrm{z \gtrsim 3}$. The results confirm the efficacy of red colour criteria for pinpointing HST-dark galaxies at $\mathrm{z>3}$, aligning with prior research that utilized NIRCam photometry \citep[e.g.][]{Barrufet2023, Gottumukkala2024, PerezGonzalez2023, Rodighiero2023, Williams2023}.

\begin{table*}
\centering
\begin{tabular}{|c|c|c|c|c|c|c|c|}
\hline
ID & RA & Dec & F$_{H\alpha}$ & F$_{H\beta}$ & Av  & EW$_{H\alpha}$ & zspec  \\
 & [deg] & [deg] & $\mathrm{[10^{-20}erg/s/cm^{2}]}$  & $\mathrm{[10^{-20}erg/s/cm^{2}]}$ & [mag]  &  $\mathrm{[\AA]}$ &  \\ 
 \hline
1260 & 53.0748575 & -27.875898   & 290   $\pm$  82   & <68           & >1.40                    & 68  $\pm$ 26     & 4.44  \\
1548 & 53.1347898 & -27.907492   & 463   $\pm$ 44    & <72           & >2.77                    & 70  $\pm$ 9      & 4.43  \\
1616 & 53.0766784 & -27.8734698  & 3553  $\pm$ 104   & <452          & >3.49                    &  331 $\pm$ 17     & 3.46    \\
3012 & 53.0767591 & -27.8641199  & 414   $\pm$ 50    & <115          & >0.80                    &   149 $\pm$ 44  & 3.47  \\
3339 & 53.1147448 & -27.8904202  & 1495  $\pm$ 50    & 230 $\pm$ 31  & 2.83                     & 304 $\pm$ 59           & 5.41 \\
3640 & 53.0821208 & -27.8598697  & 720   $\pm$ 94    & <78           & >4.03                    & 146 $\pm$ 31        & 3.65    \\
4307 & 53.1406347 & -27.8813756  & 1137  $\pm$ 55    & 83  $\pm$ 26  & 5.40                     & 375 $\pm$ 231          & 5.52 \\
4490 & 53.0410546 & -27.8544706  & 1829  $\pm$ 72    & 290 $\pm$ 42  & 2.74                     &   484 $\pm$ 81        & 3.70 \\
4820 & 53.034904  & -27.8521791  & 1966 $\pm$ 46     & 225 $\pm$ 36  & 3.85                     &   545 $\pm$ 103          & 3.79 \\
5446 & 53.0279874 & -27.8470862  & 1261  $\pm$ 51    & <209          & >2.58                    & 267 $\pm$ 24      & 2.71 \\
5510 & 53.1783286 & -27.8702734  & 1369  $\pm$ 49    & <264          & >2.06                    & 134 $\pm$ 6       & 3.47 \\
5607 & 53.1233683 & -27.8705379  & 434   $\pm$ 28    & 72  $\pm$ 28  & 2.58                     & 346 $\pm$ 38      & 3.59 \\
6151 & 53.0420756 & -27.8426437  & 316   $\pm$ 36    & <51  & >2.65  & 2025 $\pm$ 176           & 4.38 \\
12469 & 53.0446935 & -27.8136318 & 2177 $\pm$ 50     & 323 $\pm$ 63  & 2.96                     &  280 $\pm$ 17           & 3.55 \\
12577 & 53.0484576 & -27.8151421 & 1799 $\pm$ 66     & 146 $\pm$ 19  & 5.05                     &  977 $\pm$ 67          & 5.23 \\
7151  &  53.05672 &  -27.836383  & 1427 $\pm$ 67     & <119          & >4.8 & 2647 $\pm$ 93     &  3.79 \\

\\ \hline
\end{tabular}
\caption{Spectral Properties of dusty HST-Dark Galaxies. The 1st column lists the galaxy IDs; the 2nd and 3rd columns display the coordinates, RA and Dec, respectively. The 4th and 5th columns present the fluxes of $\mathrm{H\alpha}$ and $\mathrm{H\beta}$, along with their uncertainties (>$\mathrm{H\beta}$ flux indicates a 2$\sigma$ upper limit). The 6th column details the attenuation derived from the Balmer decrement (> indicates a lower limit). 
The 7th column shows the rest-frame EWs of the $\mathrm{H\alpha}$ lines. The final column contains the spectroscopic redshifts. Notice that the  $\mathrm{H\alpha}$ values are blended with [NII]. }
\label{TabSpectral_features}
\end{table*}

\subsection{Are dusty HST-dark galaxies a homogenous population?} 

\begin{figure}
 \centering
  \includegraphics[width=\columnwidth]{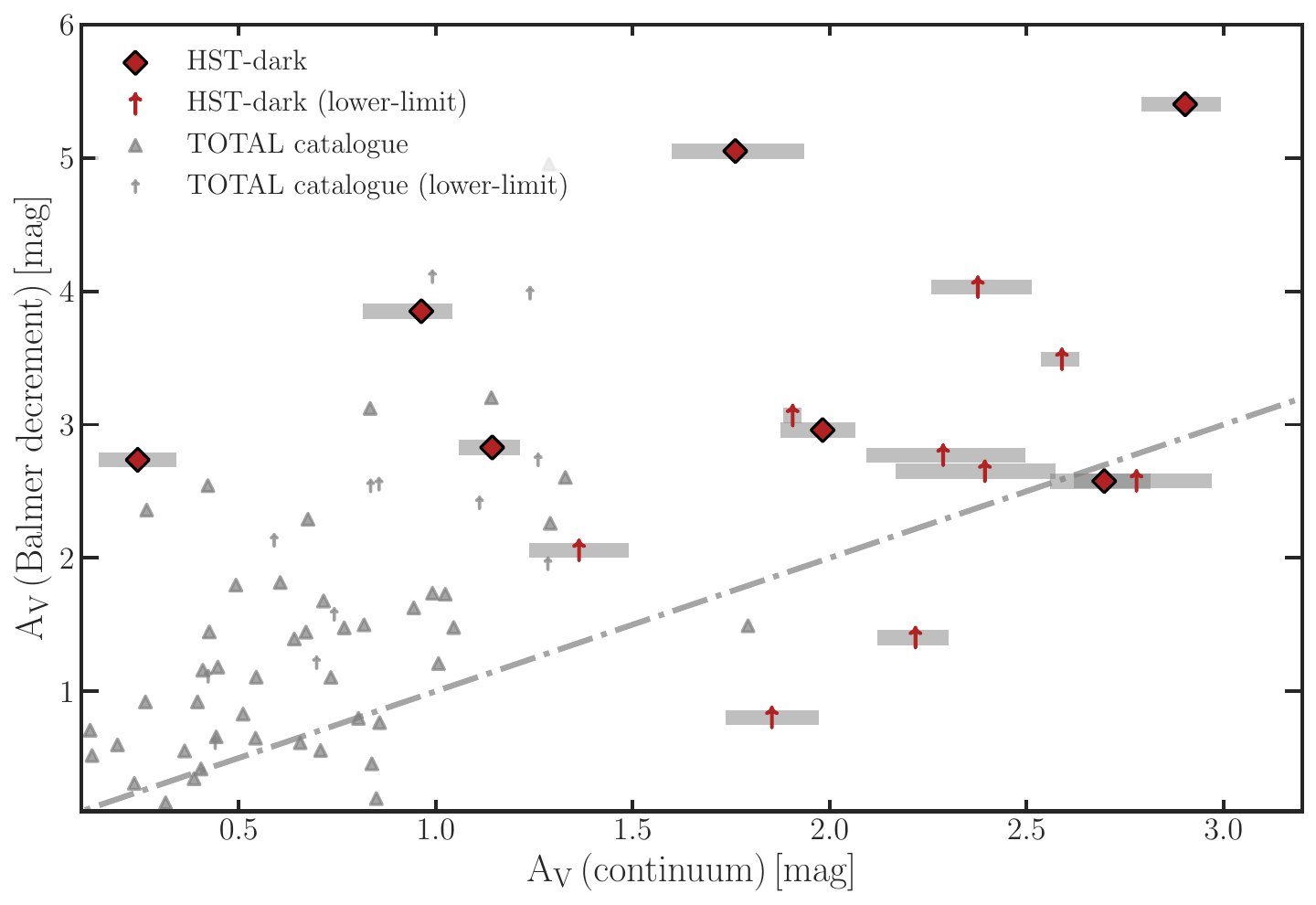}    
  \vspace{-5mm}
 \caption{Dust attenuation comparison: Balmer decrement-derived $\mathrm{A_{V}}$ versus continuum-derived $\mathrm{A_{V}}$. HST-dark galaxies that are indicated by red diamonds (with $\mathrm{H\beta}$ detection) and red arrows (2$\mathrm{\sigma}$ lower limit for non-detection; uncertainties shown by grey bars indicate the 16th and 84th percentiles) exhibit larger attenuation than the rest of the sample (non-HST dark galaxies, grey triangles and arrows). The dashed line presents the one-to-one relation. HST-dark galaxies generally surpass $\mathrm{A_{V}> 3mag}$ and even exceed $\mathrm{A_{V}> 5 mag}$. While a correlation exists between spectral and continuum $\mathrm{A_{V}}$ magnitudes, spectral measurements are consistently higher than continuum. Not surprisingly, while spectral lines probe more dust-obscured regions than continuum, there could have been a potential underestimation of dust presence in HST-dark galaxies by using photometric observations.
  }  
   \label{Balmer_decrement_Avcontinum}
\end{figure}

Despite indications of significant dust obscuration from NIRCam studies, Balmer decrements are a much more direct way of estimating dust attenuation in galaxies.
Figure \ref{Balmer_decrement_Avcontinum} compares HST-dark galaxies to the full sample of galaxies, for which we have obtained NIRSpec/PRISM spectra. HST-dark galaxies are among the most dust-enriched in the sample. Importantly, the attenuation measured from the Balmer decrement presents significantly larger values than that derived from SED fitting, with differences that vary significantly from galaxy to galaxy.
This can be expected due to the inhomogeneous distribution of gas and dust \citep[see, e.g.,][]{Calzetti2000}.

Table \ref{TabSpectral_features} presents the spectral parameters of HST-dark galaxies, such as fluxes and dust attenuation, along with $\mathrm{H\alpha}$ EWs. The E(B-V) ranges within $0.2$ to $1.3$ which translates to dust attenuations of $\mathrm{0.8 < A_{V}/mag < 5.4}$ assuming 
$\mathrm{R_{v} = 4.05}$ \citep{Kashino2013}. 
Among the galaxies with $\mathrm{H\beta}$ detections, meaning that the attenuation is computed with the $\mathrm{H\beta}$ flux and not a lower limit, the dust attenuation from Balmer lines is $\mathrm{A_{V}/mag \gtrsim 2.6}$, indicating high attenuation in HST-dark galaxies. 
Compared to the rest of the galaxy sample in our mask, HST-dark galaxies exhibit larger dust attenuation, as evidenced by both the continuum and the Balmer decrement. 

\begin{figure}
 \centering
     \includegraphics[width=\columnwidth]{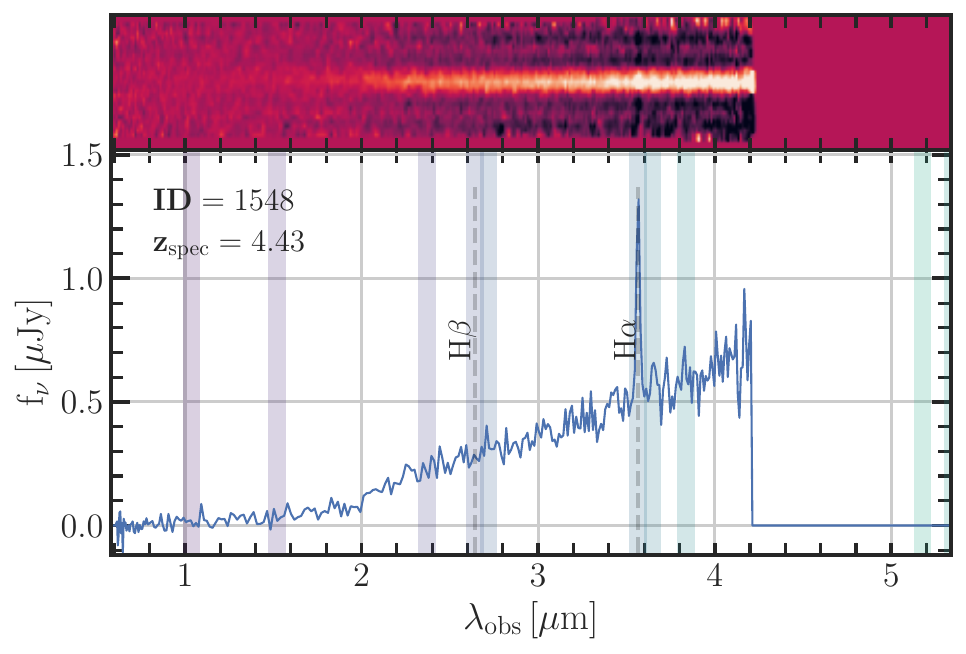} 
     \includegraphics[width=\columnwidth]{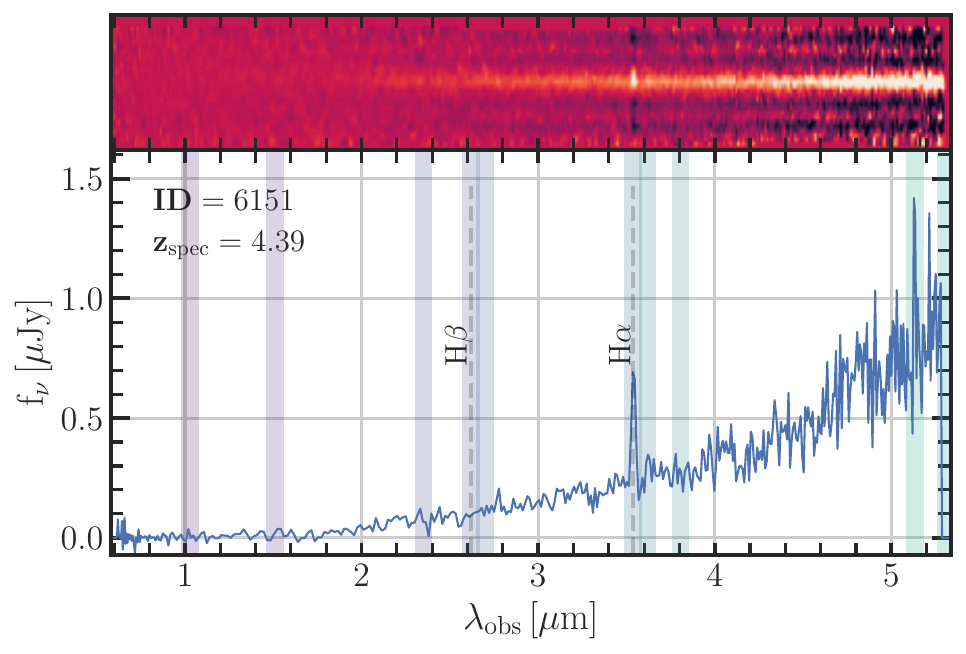}  
     \includegraphics[width=\columnwidth]{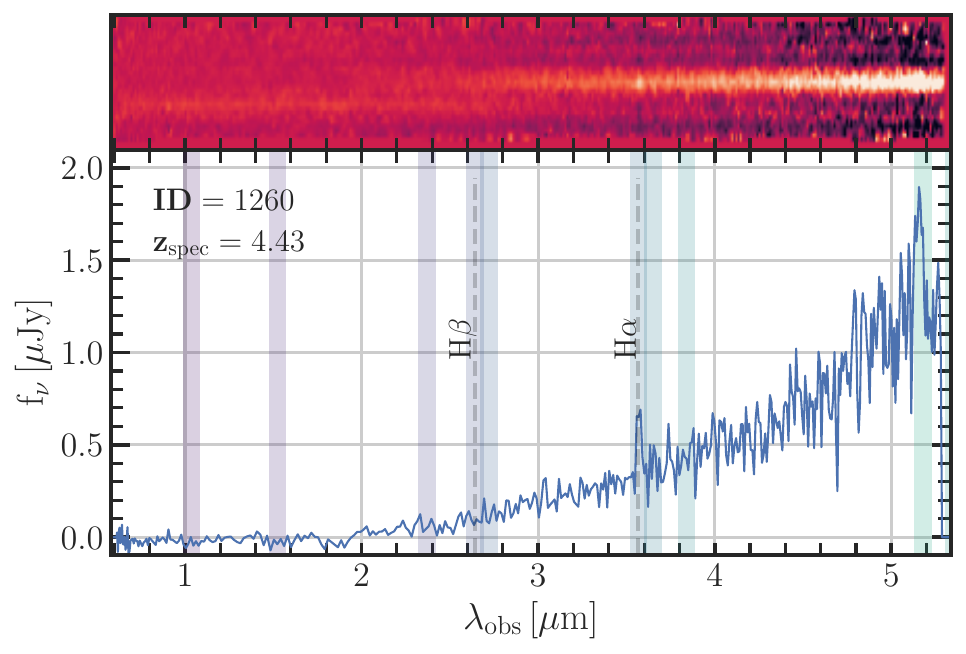}  
\caption{2D and 1D spectra from three HST-dark galaxies with extreme red slopes. The dashed lines mark the locations of the $\mathrm{H\alpha}$ and $\mathrm{H\beta}$ spectral lines. Coloured lines represent [OIII], HeI, [CI] and [SII]. These galaxies display nearly identical spectra and lie at similar redshifts of $\mathrm{z \sim 4.4}$. They show a distinct red slope with a prominent $\mathrm{H\alpha}$ line but lack $\mathrm{H\beta}$ (dashed lines) and [OIII] lines. The adjacent [SII] line (in green) to the $\mathrm{H\alpha}$ line and the detection of [SIII] at $\mathrm{\lambda_{obs} \sim 5.2 \mu m}$ support the accuracy of the redshift measurement. 
 }  
   \label{redspectra_dusty}
\end{figure}

Intriguingly, dusty HST-dark galaxies exhibit diverse spectral shapes, with a significant majority presenting red spectra. This observation prompted us to categorize them based on spectral profiles to understand their intrinsic properties better. Figure \ref{redspectra_dusty} presents the spectra of HST-dark galaxies characterized by the reddest spectra, which align with the redder $\mathrm{H-F444}$ colours. Notice that the shape of the spectra and the redshift ($\mathrm{z \sim 4.4}$) are almost identical. IDs 1548 and 6151 exhibit substantial attenuation with $\mathrm{A_{V} > 2.2 \ mag}$ whereas ID 1260 presents $\mathrm{A_{V} > 1.1 \ mag}$. Due to the lack of $\mathrm{H\beta}$ flux measurements, we cannot conclude that the galaxies have different dust attenuation. Moreover, these galaxies rank among the most massive in our dataset, akin to quiescent galaxies at similar redshifts (see physical properties in Section \ref{physical_properties}). 
Not surprisingly, colour (F160W-F444W), while indicative of mass, does not effectively distinguish galaxy nature, such as dustiness, as discussed in Section \ref{Discussion}.

\begin{figure}
 \centering
     \includegraphics[width=\columnwidth]{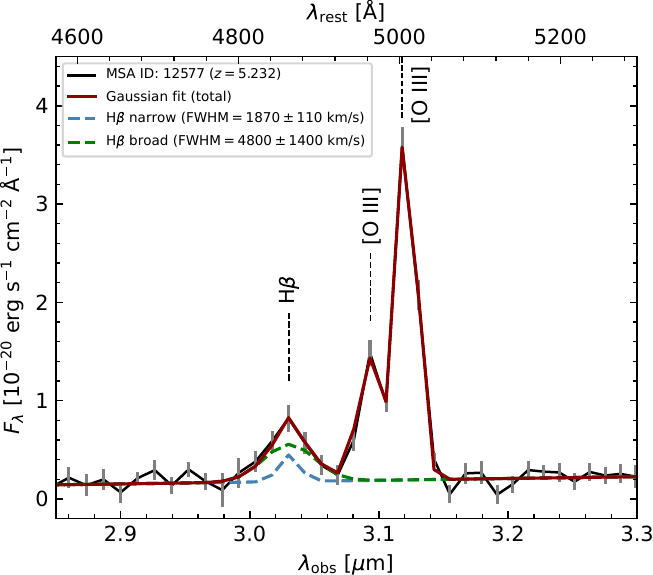}    
     \vspace{-3mm}
 \caption{Spectral fitting for the ID=12577, the only source in the sample that shows another broad line indicating strong evidence of an AGN. The black line is the spectra, whereas the red lines show the fitting for $\mathrm{H\beta}$, [OIII] and the continuum. The narrow component of the $\mathrm{H\beta}$ line is shown in blue, and the broad component is in green. 
 }  
   \label{12577_fittedspectra}
\end{figure}

\begin{figure}
 \centering    
 \includegraphics[width=\columnwidth]{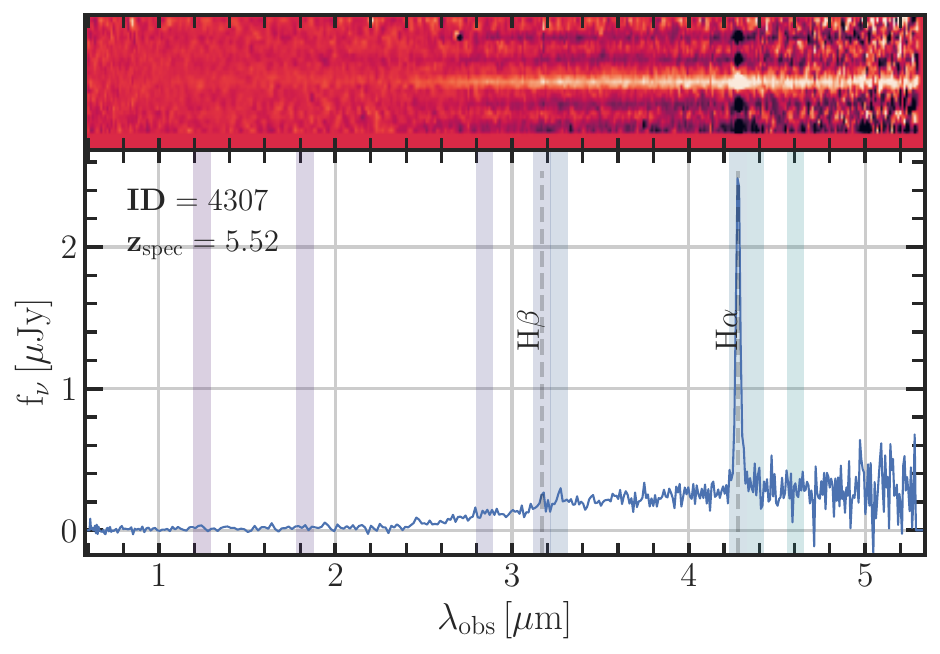} 
 \includegraphics[width=\columnwidth]{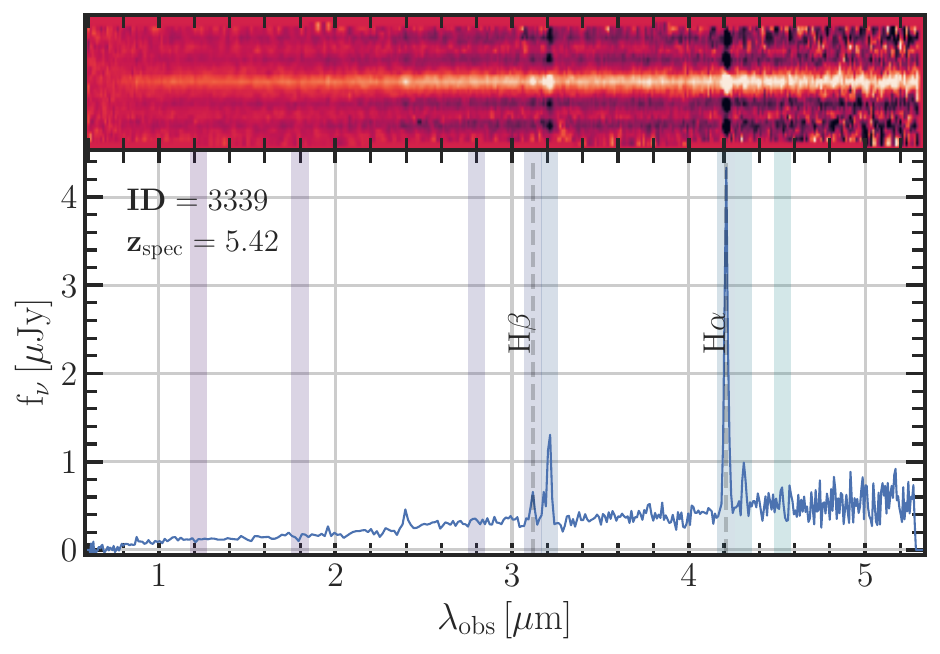} 
  \includegraphics[width=\columnwidth]{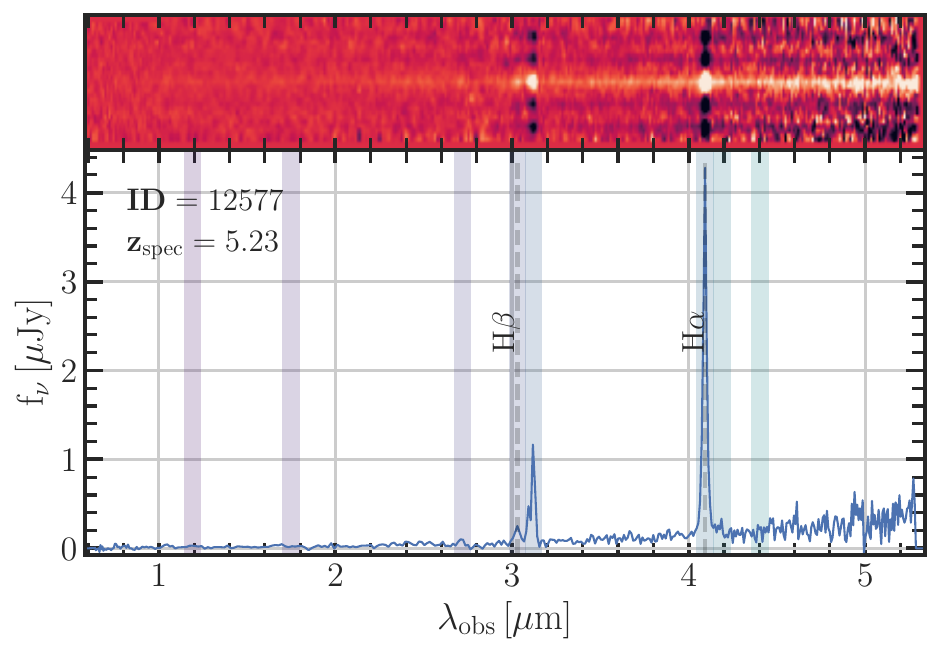} 
 
 \caption{The spectra of the three highest redshift HST-dark galaxies in our sample, with redshifts of $\mathrm{z = 5.52}$, $\mathrm{5.41}$, and $\mathrm{5.23}$. They exhibit flat spectra featuring pronounced $\mathrm{H\alpha}$, $\mathrm{H\beta}$, and [OIII] emission lines. These spectral characteristics enable precise determination of dust attenuation via the Balmer decrement, circumventing the reliance on upper limits. 
 }  
   \label{spectra_dusty_flat}
\end{figure}

In contrast to HST-dark galaxies with red-sloped spectra, Figure \ref{spectra_dusty_flat} shows HST-dark galaxies with flatter spectra and $\mathrm{H\beta}$ emission, enabling direct dust attenuation calculations. Among the highest-redshift galaxies in our sample ($\mathrm{z > 5}$), these three exhibit varying dust attenuation levels. Specifically, IDs 4307 and 12577 show substantial attenuation at $\mathrm{A_{V}/mag = 5.4}$ and $\mathrm{5.1}$, respectively, in contrast to the less severe $\mathrm{A_{V}/mag = 2.8}$ observed in ID 3339. Despite being dusty, these galaxies have moderately lower stellar masses ($\mathrm{log(M_{*}/M_{\odot}) \sim 10}$), $\mathrm{0.5 \ dex}$ below the redder HST-dark galaxies (see Section \ref{physical_properties}). We note that only source ID=12577 presents an $\mathrm{H \beta}$ line broader than $\mathrm{[OIII]}$ which suggests the presence of an AGN (see Figure \ref{12577_fittedspectra}). 

Since all dusty HST-dark galaxies are H$\alpha$ emitters, we also analysed the rest-frame EW for the $\mathrm{H\alpha}$ lines. We find a broad range of EW with $\mathrm{ 68  < EW/ \AA < 545 }$ except for the ID = 12577 with an $\mathrm{EW \sim 1000 \AA}$. Assuming the blended $\mathrm{H\alpha}$+[NII] emission predominantly originates from HII regions, with [NII] contributing minimally (but in AGN such a contribution could be notable), the inferred EWs indicate substantial recent star formation activity (within the last $\mathrm{ \sim 10Myr}$). However, it is critical to note that our observations reveal substantial variations in attenuation levels between the nebular and the stellar continuum emission (see Figure \ref{Balmer_decrement_Avcontinum}). This disparity introduces significant uncertainties in deducing star formation histories (SFHs) based solely on EW  H$\alpha$ measurements.
 
Our findings conclusively establish the dusty nature of most HST-dark galaxies through spectroscopic evidence, aligning with observations from NIRCam dust continuum data. 
While most HST-dark galaxies do not show clear AGN features, we only have evidence for one source to harbour an AGN.

\subsection{Quantifying the Quiescent Fraction in HST-Dark Galaxies}
\label{quiescentsection}

Recent JWST NIRCam results suggest that HST-dark galaxies might not only be dusty star-forming galaxies but could also exhibit quiescent characteristics \citep{GomezGuijarro2023}. Notably, \citet{PerezGonzalez2023} estimated that quiescent galaxies may represent about $\mathrm{18 \%}$ of the HST-dark/faint galaxy population. This section delves into identifying and analysing quiescent galaxies within our HST-dark galaxy sample.

\begin{figure}
 \centering
  \includegraphics[width=\columnwidth]{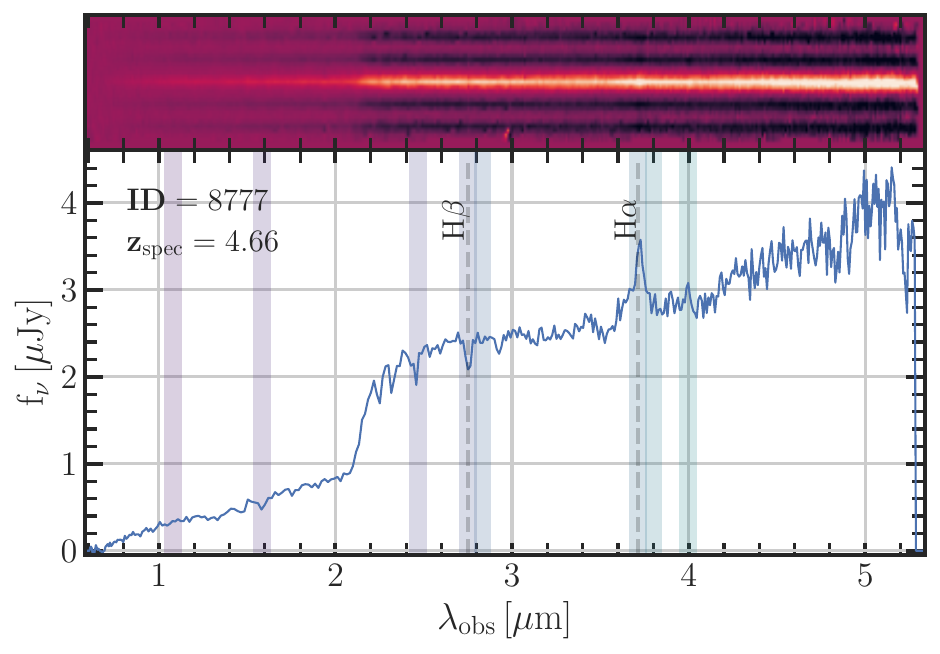 }  
     \includegraphics[width=\columnwidth]{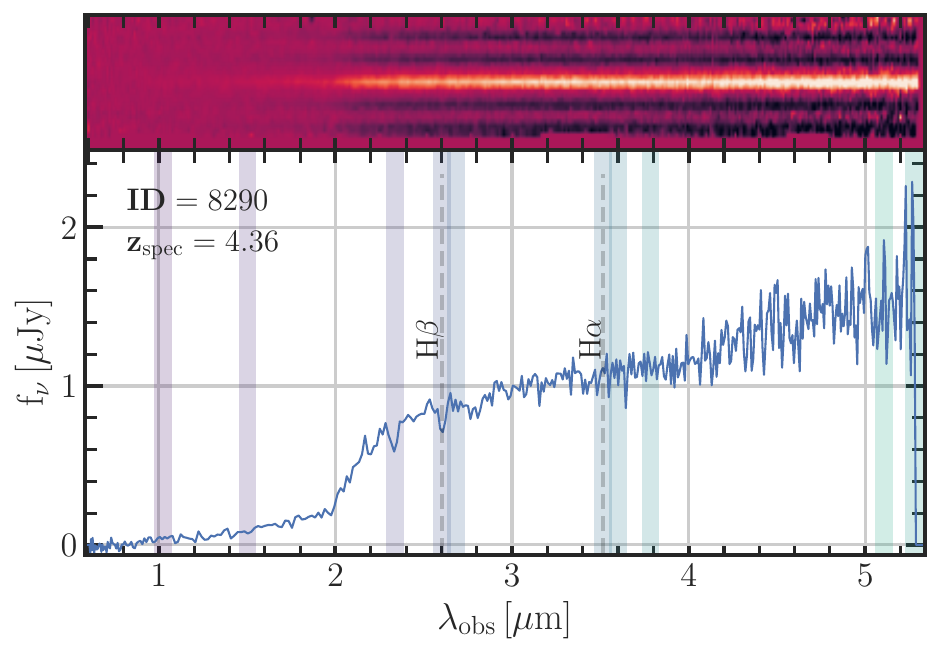}    
     \includegraphics[width=\columnwidth]{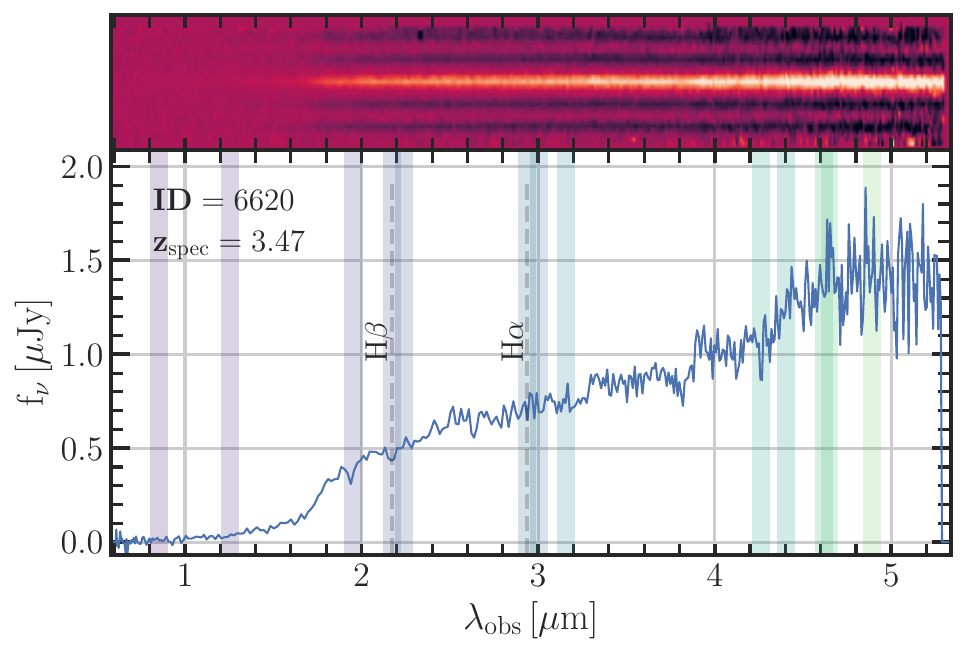}  
       \vspace{-5mm}
 \caption{2d and 1D spectra of the three quiescent galaxies of our HST-dark galaxy sample at $\mathrm{z_{spec} = 4.7, \ 4.47, \ 3.6}$ respectively. The spectra show no emission lines (dashed lines show $\mathrm{H\alpha}$ and $\mathrm{H\beta}$, and the coloured lines represent [OIII], [SII], HeI and [CI], respectively). The first spectra likely present a broad $\mathrm{H\alpha}$ line (which is blended with [NII]), whereas the other two lack this property. These quiescent galaxies, making up $\mathrm{13^{+9}_{-6} \%}$ of HST-dark galaxies in our sample, underscore the presence of inactive galaxies within the dusty HST-dark population at $\mathrm{z>3}$.}  
   \label{spectra_quiescent}
\end{figure}

Our pilot study reveals that 3 out of the 23 HST-dark galaxies, i.e., $\mathrm{13^{+9}_{-6} \%}$ are quiescent. Figure \ref{spectra_quiescent} presents the spectra of these three sources. 
The first spectrum (ID=8777) at $\mathrm{z=4.7}$ shows the highest-redshift quiescent galaxy reported to date by \citet{Carnall2023}. The presence of a broad $\mathrm{H\alpha}$ line with an observed $\mathrm{FWHM = 5720 \pm 654 km/s}$  indicates AGN activity, 
despite the low equivalent width ($\mathrm{EW = 32 \pm 5 \AA}$). 

The second quiescent galaxy (ID=8290) is at a marginally lower redshift of $\mathrm{z=4.4}$, in agreement with the photometric redshift reported by \citet{Carnall2020}. The third spectrum (ID=6620) uncovers a $\mathrm{z=3.5}$ galaxy devoid of emission lines, showing that it is a quiescent galaxy. The Balmer break is less strong than in the other two quiescent galaxies, which suggests a somewhat less older stellar population. 

Despite the limited sample size precluding definitive conclusions on galaxy evolution, the spectroscopic identification of three quiescent galaxies at $z>3$, one exhibiting clear AGN activity, raises questions about diverse quenching mechanisms. PRISM spectra suffice for redshift determination and AGN detection via the H$\mathrm{\alpha}$ line in these galaxies. However, these results underscore the need for larger and higher-resolution spectroscopic surveys to provide deeper insights into the quenching processes in galaxies of this kind. 

\subsection{Morphology of HST-dark galaxies}

This section explores the varied morphologies of HST-dark galaxies, linking their morphological features with spectral characteristics, a connection further elaborated in forthcoming work by Baggen et al. (in preparation).

\begin{figure}
  \centering
   \includegraphics[width=1\columnwidth]{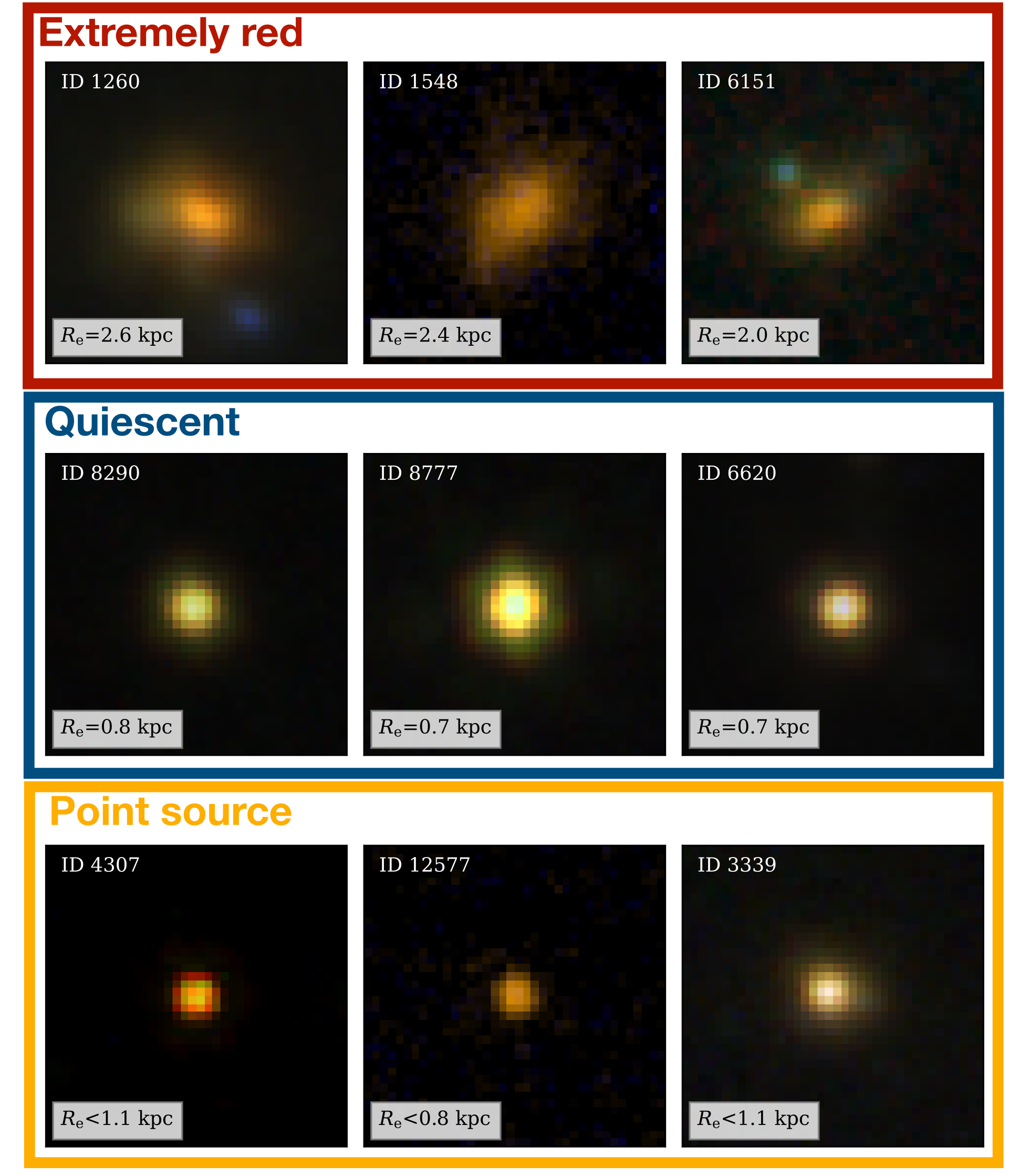}
\caption{RGB images of main HST-dark galaxies subsamples generated using the F200W, F356W, and F444W filters (blue, green, and red colours, respectively). 
The first row features extremely red galaxies (IDs: 6151, 1260, 1548), the second showcases quiescent galaxies (IDs: 8290, 8777, 6620), and the third displays the highest redshift of our sample classified as flat spectra, one of them showing broad lines and classified as AGN (IDs: 4307, 3339, 125777). HST-dark galaxies present distinct morphologies; the extremely red galaxies are more extended, with an effective radius of approximately $\mathrm{r_{e} \sim 2.5 kpc}$, in contrast to the quiescent galaxies, which are more compact with an effective radius around $\mathrm{r_{e} \sim 0.7 kpc}$. The point sources present effective radius constrained up to $\mathrm{r_{e} \lesssim 1 kpc}$.
} 
   \label{morphology}
\end{figure}

Figure \ref{morphology} showcases the RGB images within the main HST-dark galaxy subsample, categorized based on spectral analysis into extremely red galaxies, quiescent galaxies, and flat spectra (Figures \ref{redspectra_dusty}, \ref{spectra_quiescent}, \ref{spectra_dusty_flat}). 
RGB images were generated using the F200W, F356W, and F444W filters as blue, green, and red channels. For IDs 1548 and 12577, lacking F356W coverage, the F444W images were adjusted to match the F356W flux levels expected from their SED fits. The F444W images were analyzed with \textsc{galfit} to fit S\'{e}rsic profiles, characterized by a S\'{e}rsic index $n$ and an effective radius $R_\mathrm{e}$ (see Baggen et al., in preparation, for methodology details).
 
The galaxies with redder spectra typically present more expansive structures, with effective radii exceeding $\mathrm{2 \, kpc}$, which is over twice the size of the quiescent galaxies in our dataset, whose effective radii range from $\mathrm{0.66 < r_{e}/kpc < 0.76}$. Morphologically, the three quiescent galaxies exhibit a consistent structure, with an effective radius of $\mathrm{r_{e} \sim 0.7 \, \mathrm{kpc}}$ in the F444W filter. Notably, galaxy ID=8777 contains a central AGN, yet its effective radius remains in line with the other quiescent galaxies in our sample. In contrast, galaxies with $\mathrm{z_{spec} > 5}$, one of them confirmed as an AGN, manifest as compact point sources. 
We find that the effective radii are constrained to $\mathrm{r_{e} < 1 kpc}$. Despite the limited sample size, this result suggests that morphology can offer insights into the underlying nature of HST-dark galaxies (further details in Baggen et al. in preparation.).

\section{Photometric results: Are HST-dark galaxies exceptionally massive? } 
\label{physical_properties}

This section presents the physical properties of 19 HST-dark galaxies, including star formation rates (SFR), stellar masses, and dust attenuation. These parameters were derived using the photometric catalogue detailed in Section \ref{observations}. We further incorporate spectroscopic redshift data as a pivotal input, which allows us to avoid degeneracies and obtain better physical parameter constraints (see Section \ref{methodology} for methodology details). The analysis encompasses both quiescent and dust-enshrouded galaxies alongside bluer galaxies that deviate from the established red colour criterion of HST-dark galaxies.  
This comparison enables us to juxtapose HST-dark galaxies with a spectroscopic-matched sample from identical observations. 

\begin{figure*}
 \centering
     \includegraphics[width=1.7\columnwidth]{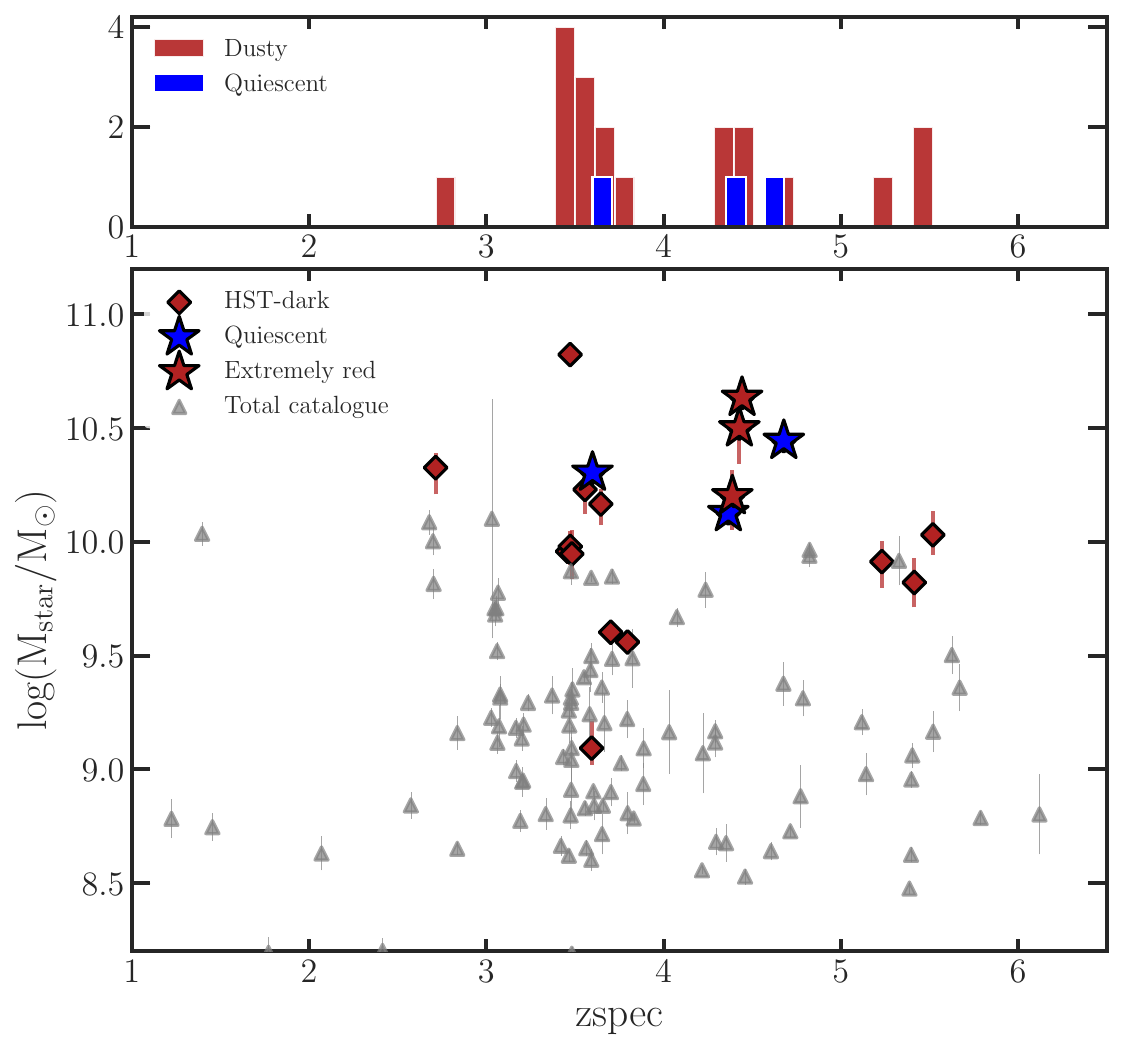}    
\caption{Spectroscopic redshift and stellar mass distribution of HST-dark galaxies. The upper diagram displays a histogram of the spectroscopic redshift for dusty (in red) and quiescent (in blue) galaxies. The lower plot displays the stellar mass vs the spectroscopic redshift, with HST-dark galaxies represented by red diamonds (general HST-dark sample). The red stars show the most extreme red HST-dark galaxies, and the blue stars show the quiescent galaxies. The remainder of the sample, the fillers in the ID 2198 Cycle 1 program, is indicated by grey triangles. HST-dark galaxies primarily lie at $\mathrm{z_{spec} > 3}$ and are among the most massive galaxies in the sample. Interestingly, the three reddest HST-dark galaxies have similar redshifts and stellar masses compared to the most distant quiescent galaxies.} 
   \label{Mstar_zspec}
\end{figure*}

In Figure \ref{Mstar_zspec}, we present the spectroscopic redshift distribution of HST-dark galaxies versus their stellar mass. HST-dark galaxies are located at $\mathrm{z_{spec} = 4.1 \pm 0.7}$, as detailed in Section \ref{Spectra_section}. These galaxies emerge as the most massive in our samples, with stellar masses ranging from $\mathrm{9.1 < log(M_{*}/M_{\odot}) < 10.8}$ with a mean of $\mathrm{log(M_{*}/M_{\odot}) = 10.1 \pm 0.4}$. Notably, the redder galaxies, which have masses around $\mathrm{log(M_{*}/M_{\odot}) = 10.5 \pm 0.2}$, are comparable to quiescent galaxies at similar redshifts ($\mathrm{z_{spec} \sim 4.4}$). The highest redshift galaxies at $\mathrm{z_{spec} > 5}$ present more moderate stellar masses, $\mathrm{log(M_{*}/M_{\odot}) \sim 10}$.

HST-dark galaxies exhibit a broad range in SFRs, with a mean of $\mathrm{SFR = 59 \pm 44 \ M_{\odot}/yr}$, yet the dustier subset consistently shows $\mathrm{SFR < 140 M_{\odot}/yr}$. This is consistent with the broad range of equivalent widths ($\mathrm{70 \AA < EW < 550 \AA}$) reported in the spectral analysis (refer to Section \ref{Spectra_section}). This indicates that HST-dark galaxies may exhibit varied star formation activities.
Specifically, sources with high equivalent widths ($\mathrm{EW \sim 500 \AA}$), such as IDs 4820 and 4490, suggest starburst activity. Yet, the average $\mathrm{EW \sim 300 \AA}$ across our sample aligns with the main sequence (MS) of galaxies \citep{Marmol-Queralto2016}. Spectral energy distribution analysis further supports this, indicating moderate SFRs and significant stellar masses, consistent with MS characteristics rather than starburst classification. It is important to note that although this study incorporates total star formation rates from BAGPIPES, it does not include far-infrared/submillimeter photometry, which may affect the SFR estimations obtained from SED fitting. Nonetheless, \cite{Williams2023} demonstrated that omitting ALMA data does not significantly alter SFRs for a similar source selection at $\mathrm{z < 5}$. 

\begin{figure*}
  \includegraphics[width=1.7\columnwidth]{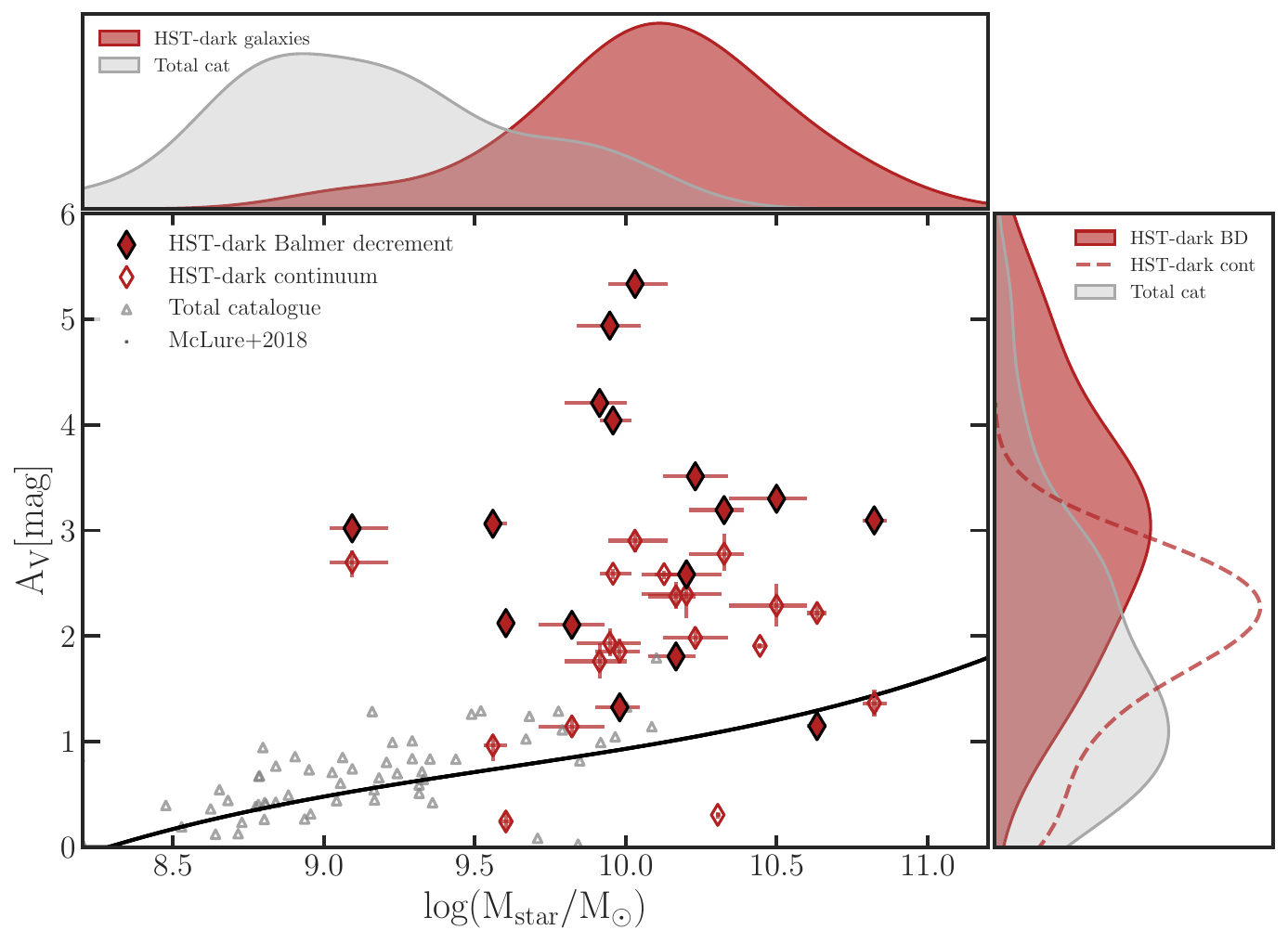}
\caption{Comparison of stellar mass and dust attenuation ($\mathrm{A_{v}}$) for HST-dark galaxies and bluer, non-HST-dark counterparts. Open symbols denote dust attenuation inferred from the continuum, whereas solid symbols indicate values obtained through Balmer decrement analysis. The black dots represent the correlation from \citet{McLure2018}, aligning with the bluer galaxies in our sample. Kernel density estimation (KDE) diagrams adjacent to the main plot (top and right) contrast the distribution of HST-dark galaxies (in red) with that of the broader sample (in grey). Attenuation values obtained through the Balmer decrement consistently exceed those derived from the continuum, exhibiting a steeper correlation.
} 
   \label{Mstar_Av}
\end{figure*}

For the dusty galaxies, we explore the well-documented correlation between stellar mass and dust attenuation in dusty galaxies \citep{Genzel2015, Whitaker2017, Fudamoto2020}. While the interdependence of these parameters is established, some research suggests this relationship may not evolve beyond $\mathrm{z > 2}$ \citep{McLure2018}. In addition, recent findings by \citet{GomezGuijarro2023} confirm a $\mathrm{M_{star}}$-$\mathrm{A_{v}}$ correlation specifically for HST-dark galaxies finding a different relation between HST-dark galaxies and Lyman Break Galaxies. Our analysis extends this work by quantifying dust attenuation for HST-dark galaxies and bluer, less massive counterparts at $\mathrm{z > 3}$, and uniquely, we compute dust attenuation using both continuum measurements and the Balmer decrement for both cases. Figure \ref{Mstar_Av} juxtaposes the stellar masses and dust attenuation of both HST-dark galaxies and also the bluer galaxies that do not satisfy the red colour criterion ($\mathrm{H-F444W < 1.75}$). Our analysis indicates that HST-dark galaxies are more massive and dustier than their bluer counterparts. HST-dark galaxies exhibit a different mass-dust attenuation correlation than the bluer galaxies, which follow a similar trend to \citep{McLure2018}, for both SED-based dust attenuation measurements and nebular emission. 
The dust attenuation values derived from the Balmer decrement exceed those obtained from the continuum (refer to Section \ref{Spectra_section}) and suggest an alternate correlation with stellar mass. 

In summary, our results confirm that HST-dark galaxies are uniquely massive and dust-rich, differing significantly from their bluer, less massive counterparts. However, we do not find that HST-dark galaxies have masses of extremely massive galaxies exceeding $\mathrm{10^{11} M_{\odot}}$. 

\section{Discussion: Is It Time to Reevaluate Red Colour Selections in the JWST Era?} 
\label{Discussion}

This section discusses the nature of HST-dark galaxies, contrasting our results with those of earlier studies. We also explain our study's caveats and propose future improvements.

As a first caveat, we notice that the quiescent subset within HST-dark galaxies exhibits considerable uncertainty ($\mathrm{13 ^{+9}_{-6} \%}$), and a larger sample would be needed to improve statistics. In addition, notice that there are biases introduced by the NIRSpec mask weighting that should warrant consideration. Utilizing NIRCam data, \citet{PerezGonzalez2023} proposed a `triality' in HST-dark galaxies, identifying $\mathrm{\sim 18 \%}$ as quiescent. While this estimate is consistent with our observations, we consider it an upper limit and advocate for more extensive spectroscopic data. Despite the limited size of our sample precluding definitive conclusions about the UVJ diagrams' reliability, our study's highest redshift quiescent galaxy is categorized within the star-forming region according to \citet{Carnall2020}. Although UVJ diagrams are extendedly used for classifying quiescent galaxies \citep{Williams2014, Valentino2020, Carnall2023, Valentino2023}, this discrepancy highlights the need to investigate alternative methods to UVJ diagrams for identifying quiescent galaxies, as suggested in the works of \citet{Carnall2018, Carnall2020, Carnall2023}.  

Following the nature of HST-dark galaxies, we find that most are dusty but have a significant disparity between spectra. Some galaxies (IDs= 6151, 1548, 1260; red stars along the paper) have extremely red spectra. These three galaxies all lie at $\mathrm{z_{spec} \sim 4.4}$, yet we discount the possibility of an overdensity at this redshift based on their spatial distribution and previously identified overdensities in GOODS-S at higher redshifts \citep{Helton2023}.  
The pronounced redness does not directly correlate with dust attenuation, as these galaxies exhibit varied levels of dust attenuation. In a related finding, \citet{ArrabalHaro2023} presented a galaxy with a similar red spectrum at $\mathrm{z_{spec} = 4.9}$, initially reported in \citet{Barrufet2023} and further discussed by \citet{Zavala2023}. This pattern may indicate a selection bias wherein redder colours are more likely to identify the most massive HST-dark galaxies, akin to the `Spitzer HST-dark' galaxies described by \citet{Wang2019}. 

\citet{PerezGonzalez2024} identified highly attenuated sources ($\mathrm{A_{v} \sim 10 \ mag}$) at $\mathrm{z_{spec} > 5}$, suggesting a combination of star formation and AGN activity in compact HST-dark galaxies, the so-called `little red dots'. Conversely, \citet{Matthee2023b, Kocevski2023, Greene2023} supports that these compact sources are AGN. Furthermore, the LRD definition has evolved from its original selection criteria, with various authors adopting disparate definitions for these sources. Our data indicate at least one compact point source at $\mathrm{z_{spec} > 5}$ is an AGN, with lower attenuation levels than \citet{PerezGonzalez2024} ($\mathrm{A_{v} \lesssim 5 \ mag}$). We also highlight the issue of blending $\mathrm{H\alpha+[NII]}$ in low-resolution NIRSpec spectra, which could inflate dust attenuation estimates, potentially reducing actual attenuation to $\mathrm{A_{v} \sim 2 \ mag}$ if [NII] flux contributes significantly (up to 40 \%). 

The discovery of large stellar masses in HST-dark galaxies by initial JWST NIRCam studies \citep{Labbe2023a} and subsequent confirmation with FRESCO data \citep{Xiao2023} have sparked considerable discussion. Contrarily, some works suggest that incorporating MIRI data might lead to lower stellar mass estimates \citep{Williams2023, PerezGonzalez2024}, a point recently emphasized by \citep{Wang2024}. Moreover, \citet{Barro2023} revealed that stellar mass estimates could vary by approximately $\mathrm{1 \ dex}$ based on the choice of SED fitting code. In a distinct approach, \citet{Lu2024} reassessed stellar population models, initially tested on a limited set of galaxies, raising the question of their potential impact on the mass estimates of red galaxies. Although various factors could influence stellar mass determinations, extending beyond this paper's focus, our analysis indicates that our sample of HST-dark galaxies at $\mathrm{z_{spec} > 5}$ do not exhibit anomalously high masses, aligning with current cosmological models.

This work demonstrates, via the Balmer decrement, that $\mathrm{\sim 90 \%}$ of HST-dark galaxies are very dusty. However, we cannot assess the total obscured star formation since the absence of far-infrared and submillimeter data limits us. To accurately determine the total SFR, it is essential to incorporate ALMA data. \citet{Wang2019} have provided observations for three galaxies within this sample. However, the rest of this sample has been unexplored so far. Concurrently, several studies show the serendipitous dust continuum detections of UV-bright galaxies \citep{Fudamoto2021, Barrufet2023b}, altogether suggesting the need for deeper observations. Future ALMA cycles will be critical in fully characterizing HST-dark galaxies.  

Our study reveals that HST-dark galaxies 
-- a term encompassing H-dropouts, red galaxies, and closely related to HST-faint, optically-faint, optically invisible galaxies -- 
span diverse characteristics, including quiescent, dusty, and potentially AGN-influenced nature, alongside variations in morphology and mass. This diversity prompts the consideration of whether, in the JWST era, we should shift towards selecting galaxies based on physical properties (i.e. mass, redshift). While such selections pose challenges, the accumulated knowledge and the integration of JWST instruments like NIRCam, NIRSpec, and MIRI in extensive multi-band surveys suggest a possible transition towards physical properties selection criteria. 

\section{Summary and Conclusions} 
\label{Summary}

Through the 10hr Cycle-1 program (ID 2198, PIs Barrufet, Oesch) titled `Quiescent or dusty? Unveiling the nature of extremely red galaxies at $\mathrm{z>3}$', we have analyzed the spectra of 23 HST-dark galaxies ($\mathrm{H-F444W>1.75}$). While these galaxies are generally presumed to be dusty, there has been speculation about the presence of quiescent galaxies among them, as photometric data alone, even from NIRCam data, cannot reliably differentiate between the two galaxy populations. Our program shows that low-resolution NIRSpec/PRISM spectra can discern between quiescent and dusty galaxies. 

The main results of the paper are:

\begin{itemize}
    \item The mean spectroscopic redshift of HST-dark galaxies is $\mathrm{z_{spec} = 4.1 \pm 0.7}$ with the entire sample lying at $\mathrm{z_{spec} \gtrsim 3}$. 
    
    \item HST-dark galaxies are mostly dusty, characterized by a dust attenuation from the Balmer decrement of $\mathrm{A_{V} = 3.1 \pm 1.2}$, exceeding the dust attenuation values obtained from continuum measurements, which stand at $\mathrm{A_{V} = 1.9 \pm 0.8}$.  

    \item Dusty HST-dark galaxies are $\mathrm{H\alpha}$ emitters with equivalent widths ranging from  $\mathrm{  68 \AA <  EW_{H\alpha} < 550 \AA }$ suggestion a wide range of recent star formation.

    \item Among HST-dark galaxies, hidden gems are discovered: quiescent galaxies at $\mathrm{z_{spec} > 3.5}$. Our program enables us to estimate that $\mathrm{13 \%}$ of HST-dark galaxies are quiescent. Particularly, the highest redshift ($\mathrm{z_{spec} = 4.7}$) passive galaxy has a broad line of $\mathrm{FWHM = 5523 km/s}$ unequivocally containing an AGN. 
    
    \item Morphological analysis of HST-dark galaxies reveals size variations correlating with spectral types: extremely red galaxies are more extended, averaging $\mathrm{R_{e} \sim 2.4 kpc}$, quiescent galaxies are compact, typically around $\mathrm{R_{e} \sim 0.7 kpc}$. The three very compact sources at $\mathrm{z_{spec}>5}$ present radius $\mathrm{R_{e} < 1 kpc}$, but we only confirm AGN in one of them. 
    
    \item HST-dark galaxies are predominantly massive with $\mathrm{\sim 85 \%}$ of the sample having stellar masses of $\mathrm{log(M_{*}/M_{\odot})> = 9.8}$.
    Notably, the extremely red galaxies are among the most massive HST-dark galaxies ($\mathrm{log(M_{*}/M_{\odot}) = 10.5 \pm 0.2}$). These masses are on par with those of the quiescent HST-dark galaxies in this sample at similar redshifts $\mathrm{ z_{spec} \sim 4.4}$.

\end{itemize}    

Our study has significantly advanced our understanding by showing that HST-dark galaxies constitute a non-homogeneous population. While colour is a reliable indicator of mass, it falls short of fully revealing the nature of HST-dark galaxies. The trend observed is that the redder galaxies are the most massive, encompassing quiescent and dusty types. 

This pilot program has effectively shown the NIRSpec/PRISM's ability to distinguish between quiescent and dusty galaxies and to shed light on dust attenuation via spectral analysis. Nonetheless, more extensive surveys with higher spectral resolution are required for precise attenuation quantification and exploration of the metallicity-mass relationship in these galaxies. Additionally, the potential synergies between JWST and ALMA are yet to be fully leveraged. While this investigation underscores NIRSpec/PRISM's capabilities, a thorough understanding of dust properties in HST-dark galaxies necessitates combined analyses with ALMA's submillimeter observations.

\section*{Acknowledgements}

The authors thank and acknowledge the effort of the program coordinator of this program, Wilson Joy Skipper, the NIRSpec reviewer, Tim Rawle, and the NIRCam reviewer, Dan Coe to make possible these observations. 
We acknowledge support from the Swiss National Science Foundation through project grant 200020\_207349 (LB, PAO, and AW). 
This work has received funding from the Swiss State Secretariat for Education, Research and Innovation (SERI) under contract number MB22.00072.
The Cosmic Dawn Center (DAWN) is funded by the Danish National Research Foundation under grant DNRF140.
YF acknowledges support from NAOJ ALMA Scientific Research Grant number 2020-16B and JSPS KAKENHI Grant Numbers JP22K21349 and JP23K13149. 
MS acknowledges support from the CIDEGENT/2021/059 grant and project PID2019-109592GB-I00/AEI/10.13039/501100011033 from the Spanish Ministerio de Ciencia e Innovacion - Agencia Estatal de Investigacion.
RJB and MS acknowledge support from NWO grant TOP1.16.057. JSD and LB thank the Royal Society for the support of a Royal Society Research Professorship. RJM and DJM acknowledge the support of the Science and Technology Facilities Council.
 M.J.M.~acknowledges the support of 
the National Science Centre, Poland through the SONATA BIS grant 2018/30/E/ST9/00208 and
the Polish National Agency for Academic Exchange Bekker grant BPN/BEK/2022/1/00110.
This work is based on observations made with the NASA/ESA/CSA James Webb Space Telescope. The raw data were obtained from the Mikulski Archive for Space Telescopes at the Space Telescope Science Institute, which is operated by the Association of Universities for Research in Astronomy, Inc., under NASA contract NAS 5-03127 for \textit{JWST}. These observations are associated with program \#2198.

Facilities: \textit{JWST}, \textit{HST}

Software:
    \texttt{matplotlib} \citep{matplotlib},
    \texttt{numpy} \citep{numpy},
    \texttt{scipy} \citep{scipy},
    \texttt{jupyter} \citep{jupyter},
    \texttt{Astropy}
    \citep{astropy1, astropy2},
    \texttt{grizli} \citep{grizli},
    \texttt{SExtractor} \citep{Bertin96},
    \texttt{bagpipes} \citep{Carnall2018}

\section*{Data Availability}

The data products are available from the authors upon reasonable request.


\bibliographystyle{mnras}
\bibliography{Hdropspaper} 

\begin{thebibliography}{}
\makeatletter
\relax
\def\mn@urlcharsother{\let\do\@makeother \do\$\do\&\do\#\do\^\do\_\do\%\do\~}
\def\mn@doi{\begingroup\mn@urlcharsother \@ifnextchar [ {\mn@doi@}
  {\mn@doi@[]}}
\def\mn@doi@[#1]#2{\def\@tempa{#1}\ifx\@tempa\@empty \href
  {http://dx.doi.org/#2} {doi:#2}\else \href {http://dx.doi.org/#2} {#1}\fi
  \endgroup}
\def\mn@eprint#1#2{\mn@eprint@#1:#2::\@nil}
\def\mn@eprint@arXiv#1{\href {http://arxiv.org/abs/#1} {{\tt arXiv:#1}}}
\def\mn@eprint@dblp#1{\href {http://dblp.uni-trier.de/rec/bibtex/#1.xml}
  {dblp:#1}}
\def\mn@eprint@#1:#2:#3:#4\@nil{\def\@tempa {#1}\def\@tempb {#2}\def\@tempc
  {#3}\ifx \@tempc \@empty \let \@tempc \@tempb \let \@tempb \@tempa \fi \ifx
  \@tempb \@empty \def\@tempb {arXiv}\fi \@ifundefined
  {mn@eprint@\@tempb}{\@tempb:\@tempc}{\expandafter \expandafter \csname
  mn@eprint@\@tempb\endcsname \expandafter{\@tempc}}}

\bibitem[\protect\citeauthoryear{{Akins} et~al.,}{{Akins}
  et~al.}{2023}]{Akins2023}
{Akins} H.~B.,  et~al., 2023, \mn@doi [arXiv e-prints]
  {10.48550/arXiv.2304.12347}, \href
  {https://ui.adsabs.harvard.edu/abs/2023arXiv230412347A} {p. arXiv:2304.12347}

\bibitem[\protect\citeauthoryear{{Alcalde Pampliega} et~al.,}{{Alcalde
  Pampliega} et~al.}{2019}]{AlcaldePampliega2019}
{Alcalde Pampliega} B.,  et~al., 2019, \mn@doi [\apj]
  {10.3847/1538-4357/ab14f2}, \href
  {https://ui.adsabs.harvard.edu/abs/2019ApJ...876..135A} {876, 135}

\bibitem[\protect\citeauthoryear{{Arrabal Haro} et~al.,}{{Arrabal Haro}
  et~al.}{2023}]{ArrabalHaro2023}
{Arrabal Haro} P.,  et~al., 2023, \mn@doi [\nat] {10.1038/s41586-023-06521-7},
  \href {https://ui.adsabs.harvard.edu/abs/2023Natur.622..707A} {622, 707}

\bibitem[\protect\citeauthoryear{{Astropy Collaboration} et~al.,}{{Astropy
  Collaboration} et~al.}{2013}]{astropy1}
{Astropy Collaboration} et~al., 2013, \mn@doi [\aap]
  {10.1051/0004-6361/201322068}, \href
  {http://adsabs.harvard.edu/abs/2013A%26A...558A..33A} {558, A33}

\bibitem[\protect\citeauthoryear{{Astropy Collaboration} et~al.,}{{Astropy
  Collaboration} et~al.}{2018}]{astropy2}
{Astropy Collaboration} et~al., 2018, \mn@doi [\aj] {10.3847/1538-3881/aabc4f},
  \href {https://ui.adsabs.harvard.edu/abs/2018AJ....156..123A} {156, 123}

\bibitem[\protect\citeauthoryear{{Barro} et~al.,}{{Barro}
  et~al.}{2023}]{Barro2023}
{Barro} G.,  et~al., 2023, \mn@doi [arXiv e-prints]
  {10.48550/arXiv.2305.14418}, \href
  {https://ui.adsabs.harvard.edu/abs/2023arXiv230514418B} {p. arXiv:2305.14418}

\bibitem[\protect\citeauthoryear{{Barrufet} et~al.,}{{Barrufet}
  et~al.}{2023a}]{Barrufet2023}
{Barrufet} L.,  et~al., 2023a, \mn@doi [\mnras] {10.1093/mnras/stad947}, \href
  {https://ui.adsabs.harvard.edu/abs/2023MNRAS.522..449B} {522, 449}

\bibitem[\protect\citeauthoryear{{Barrufet} et~al.,}{{Barrufet}
  et~al.}{2023b}]{Barrufet2023b}
{Barrufet} L.,  et~al., 2023b, \mn@doi [\mnras] {10.1093/mnras/stad1259}, \href
  {https://ui.adsabs.harvard.edu/abs/2023MNRAS.522.3926B} {522, 3926}

\bibitem[\protect\citeauthoryear{{Bertin} \& {Arnouts}}{{Bertin} \&
  {Arnouts}}{1996}]{Bertin96}
{Bertin} E.,  {Arnouts} S.,  1996, \aaps, \href
  {http://adsabs.harvard.edu/abs/1996A%26AS..117..393B} {117, 393}

\bibitem[\protect\citeauthoryear{{Boylan-Kolchin}}{{Boylan-Kolchin}}{2023}]{BoylanKolchin2023}
{Boylan-Kolchin} M.,  2023, \mn@doi [Nature Astronomy]
  {10.1038/s41550-023-01937-7}, \href
  {https://ui.adsabs.harvard.edu/abs/2023NatAs.tmp...77B} {}

\bibitem[\protect\citeauthoryear{{Brammer}}{{Brammer}}{2018}]{grizli}
{Brammer} G.,  2018, {Gbrammer/Grizli: Preliminary Release}, Zenodo,
  \mn@doi{10.5281/zenodo.1146905}

\bibitem[\protect\citeauthoryear{{Bruzual} \& {Charlot}}{{Bruzual} \&
  {Charlot}}{2003}]{Bruzual2003}
{Bruzual} G.,  {Charlot} S.,  2003, \mn@doi [\mnras]
  {10.1046/j.1365-8711.2003.06897.x}, \href
  {http://adsabs.harvard.edu/abs/2003MNRAS.344.1000B} {344, 1000}

\bibitem[\protect\citeauthoryear{{Calzetti}, {Armus}, {Bohlin}, {Kinney},
  {Koornneef}  \& {Storchi-Bergmann}}{{Calzetti} et~al.}{2000}]{Calzetti2000}
{Calzetti} D.,  {Armus} L.,  {Bohlin} R.~C.,  {Kinney} A.~L.,  {Koornneef} J.,
   {Storchi-Bergmann} T.,  2000, \mn@doi [\apj] {10.1086/308692}, \href
  {http://adsabs.harvard.edu/abs/2000ApJ...533..682C} {533, 682}

\bibitem[\protect\citeauthoryear{{Caputi} et~al.,}{{Caputi}
  et~al.}{2012}]{Caputi2012}
{Caputi} K.~I.,  et~al., 2012, \mn@doi [\apjl] {10.1088/2041-8205/750/1/L20},
  \href {https://ui.adsabs.harvard.edu/abs/2012ApJ...750L..20C} {750, L20}

\bibitem[\protect\citeauthoryear{{Carnall}, {McLure}, {Dunlop}  \&
  {Dav{\'e}}}{{Carnall} et~al.}{2018}]{Carnall2018}
{Carnall} A.~C.,  {McLure} R.~J.,  {Dunlop} J.~S.,   {Dav{\'e}} R.,  2018,
  \mn@doi [\mnras] {10.1093/mnras/sty2169}, \href
  {https://ui.adsabs.harvard.edu/abs/2018MNRAS.480.4379C} {480, 4379}

\bibitem[\protect\citeauthoryear{{Carnall} et~al.,}{{Carnall}
  et~al.}{2020}]{Carnall2020}
{Carnall} A.~C.,  et~al., 2020, \mn@doi [\mnras] {10.1093/mnras/staa1535},
  \href {https://ui.adsabs.harvard.edu/abs/2020MNRAS.496..695C} {496, 695}

\bibitem[\protect\citeauthoryear{{Carnall} et~al.,}{{Carnall}
  et~al.}{2023a}]{Carnall2023}
{Carnall} A.~C.,  et~al., 2023a, \mn@doi [arXiv e-prints]
  {10.48550/arXiv.2301.11413}, \href
  {https://ui.adsabs.harvard.edu/abs/2023arXiv230111413C} {p. arXiv:2301.11413}

\bibitem[\protect\citeauthoryear{{Carnall} et~al.,}{{Carnall}
  et~al.}{2023b}]{Carnall2023b}
{Carnall} A.~C.,  et~al., 2023b, \mn@doi [\mnras] {10.1093/mnras/stad369},
  \href {https://ui.adsabs.harvard.edu/abs/2023MNRAS.520.3974C} {520, 3974}

\bibitem[\protect\citeauthoryear{{Dom{\'\i}nguez} et~al.,}{{Dom{\'\i}nguez}
  et~al.}{2013}]{Dominguez2013}
{Dom{\'\i}nguez} A.,  et~al., 2013, \mn@doi [\apj]
  {10.1088/0004-637X/763/2/145}, \href
  {https://ui.adsabs.harvard.edu/abs/2013ApJ...763..145D} {763, 145}

\bibitem[\protect\citeauthoryear{{Eisenstein} et~al.,}{{Eisenstein}
  et~al.}{2023a}]{Eisenstein2023}
{Eisenstein} D.~J.,  et~al., 2023a, \mn@doi [arXiv e-prints]
  {10.48550/arXiv.2306.02465}, \href
  {https://ui.adsabs.harvard.edu/abs/2023arXiv230602465E} {p. arXiv:2306.02465}

\bibitem[\protect\citeauthoryear{{Eisenstein} et~al.,}{{Eisenstein}
  et~al.}{2023b}]{Eisenstein2023b}
{Eisenstein} D.~J.,  et~al., 2023b, \mn@doi [arXiv e-prints]
  {10.48550/arXiv.2310.12340}, \href
  {https://ui.adsabs.harvard.edu/abs/2023arXiv231012340E} {p. arXiv:2310.12340}

\bibitem[\protect\citeauthoryear{{Ferland} et~al.,}{{Ferland}
  et~al.}{2017}]{Ferland2017}
{Ferland} G.~J.,  et~al., 2017, \mn@doi [\rmxaa] {10.48550/arXiv.1705.10877},
  \href {https://ui.adsabs.harvard.edu/abs/2017RMxAA..53..385F} {53, 385}

\bibitem[\protect\citeauthoryear{{Feroz}, {Hobson}, {Cameron}  \&
  {Pettitt}}{{Feroz} et~al.}{2019}]{Feroz2019}
{Feroz} F.,  {Hobson} M.~P.,  {Cameron} E.,   {Pettitt} A.~N.,  2019, \mn@doi
  [The Open Journal of Astrophysics] {10.21105/astro.1306.2144}, \href
  {https://ui.adsabs.harvard.edu/abs/2019OJAp....2E..10F} {2, 10}

\bibitem[\protect\citeauthoryear{{Franco} et~al.,}{{Franco}
  et~al.}{2018}]{Franco2018}
{Franco} M.,  et~al., 2018, \mn@doi [\aap] {10.1051/0004-6361/201832928}, \href
  {https://ui.adsabs.harvard.edu/abs/2018A&A...620A.152F} {620, A152}

\bibitem[\protect\citeauthoryear{{Fudamoto} et~al.,}{{Fudamoto}
  et~al.}{2020}]{Fudamoto2020}
{Fudamoto} Y.,  et~al., 2020, \mn@doi [\aap] {10.1051/0004-6361/202038163},
  \href {https://ui.adsabs.harvard.edu/abs/2020A&A...643A...4F} {643, A4}

\bibitem[\protect\citeauthoryear{{Fudamoto} et~al.,}{{Fudamoto}
  et~al.}{2021}]{Fudamoto2021}
{Fudamoto} Y.,  et~al., 2021, \mn@doi [\nat] {10.1038/s41586-021-03846-z},
  \href {https://ui.adsabs.harvard.edu/abs/2021Natur.597..489F} {597, 489}

\bibitem[\protect\citeauthoryear{{Genzel} et~al.,}{{Genzel}
  et~al.}{2015}]{Genzel2015}
{Genzel} R.,  et~al., 2015, \mn@doi [\apj] {10.1088/0004-637X/800/1/20}, \href
  {https://ui.adsabs.harvard.edu/abs/2015ApJ...800...20G} {800, 20}

\bibitem[\protect\citeauthoryear{{G{\'o}mez-Guijarro}
  et~al.,}{{G{\'o}mez-Guijarro} et~al.}{2023}]{GomezGuijarro2023}
{G{\'o}mez-Guijarro} C.,  et~al., 2023, \mn@doi [\aap]
  {10.1051/0004-6361/202346673}, \href
  {https://ui.adsabs.harvard.edu/abs/2023A&A...677A..34G} {677, A34}

\bibitem[\protect\citeauthoryear{Gottumukkala et~al.,}{Gottumukkala
  et~al.}{2024}]{Gottumukkala2024}
Gottumukkala R.,  et~al., 2024, \mn@doi [MNRAS] {10.1093/mnras/stae754}, p.
  stae754

\bibitem[\protect\citeauthoryear{{Greene} et~al.,}{{Greene}
  et~al.}{2023}]{Greene2023}
{Greene} J.~E.,  et~al., 2023, \mn@doi [arXiv e-prints]
  {10.48550/arXiv.2309.05714}, \href
  {https://ui.adsabs.harvard.edu/abs/2023arXiv230905714G} {p. arXiv:2309.05714}

\bibitem[\protect\citeauthoryear{{Gruppioni} et~al.,}{{Gruppioni}
  et~al.}{2020}]{Gruppioni2020}
{Gruppioni} C.,  et~al., 2020, \mn@doi [\aap] {10.1051/0004-6361/202038487},
  \href {https://ui.adsabs.harvard.edu/abs/2020A&A...643A...8G} {643, A8}

\bibitem[\protect\citeauthoryear{{Helton} et~al.,}{{Helton}
  et~al.}{2023}]{Helton2023}
{Helton} J.~M.,  et~al., 2023, \mn@doi [arXiv e-prints]
  {10.48550/arXiv.2311.04270}, \href
  {https://ui.adsabs.harvard.edu/abs/2023arXiv231104270H} {p. arXiv:2311.04270}

\bibitem[\protect\citeauthoryear{Hunter}{Hunter}{2007}]{matplotlib}
Hunter J.~D.,  2007, \mn@doi [Computing In Science \& Engineering]
  {10.1109/MCSE.2007.55}, 9, 90

\bibitem[\protect\citeauthoryear{{Kashino} et~al.,}{{Kashino}
  et~al.}{2013}]{Kashino2013}
{Kashino} D.,  et~al., 2013, \mn@doi [\apjl] {10.1088/2041-8205/777/1/L8},
  \href {https://ui.adsabs.harvard.edu/abs/2013ApJ...777L...8K} {777, L8}

\bibitem[\protect\citeauthoryear{Kluyver et~al.,}{Kluyver
  et~al.}{2016}]{jupyter}
Kluyver T.,  et~al., 2016, in Loizides F.,  Schmidt B.,  eds, Positioning and
  Power in Academic Publishing: Players, Agents and Agendas. pp 87 -- 90

\bibitem[\protect\citeauthoryear{{Kocevski} et~al.,}{{Kocevski}
  et~al.}{2023}]{Kocevski2023}
{Kocevski} D.~D.,  et~al., 2023, \mn@doi [arXiv e-prints]
  {10.48550/arXiv.2302.00012}, \href
  {https://ui.adsabs.harvard.edu/abs/2023arXiv230200012K} {p. arXiv:2302.00012}

\bibitem[\protect\citeauthoryear{{Kroupa}}{{Kroupa}}{2001}]{Kroupa2001}
{Kroupa} P.,  2001, \mn@doi [\mnras] {10.1046/j.1365-8711.2001.04022.x}, \href
  {https://ui.adsabs.harvard.edu/abs/2001MNRAS.322..231K} {322, 231}

\bibitem[\protect\citeauthoryear{{Labb{\'e}} et~al.,}{{Labb{\'e}}
  et~al.}{2015}]{Labbe2015}
{Labb{\'e}} I.,  et~al., 2015, \mn@doi [\apjs] {10.1088/0067-0049/221/2/23},
  \href {https://ui.adsabs.harvard.edu/abs/2015ApJS..221...23L} {221, 23}

\bibitem[\protect\citeauthoryear{{Labb{\'e}} et~al.,}{{Labb{\'e}}
  et~al.}{2023}]{Labbe2023a}
{Labb{\'e}} I.,  et~al., 2023, \mn@doi [\nat] {10.1038/s41586-023-05786-2},
  \href {https://ui.adsabs.harvard.edu/abs/2023Natur.616..266L} {616, 266}

\bibitem[\protect\citeauthoryear{{Lovell}, {Harrison}, {Harikane}, {Tacchella}
  \& {Wilkins}}{{Lovell} et~al.}{2023}]{Lovell2023}
{Lovell} C.~C.,  {Harrison} I.,  {Harikane} Y.,  {Tacchella} S.,   {Wilkins}
  S.~M.,  2023, \mn@doi [\mnras] {10.1093/mnras/stac3224}, \href
  {https://ui.adsabs.harvard.edu/abs/2023MNRAS.518.2511L} {518, 2511}

\bibitem[\protect\citeauthoryear{{Lu} et~al.,}{{Lu} et~al.}{2024}]{Lu2024}
{Lu} S.,  et~al., 2024, \mn@doi [arXiv e-prints] {10.48550/arXiv.2403.07414},
  \href {https://ui.adsabs.harvard.edu/abs/2024arXiv240307414L} {p.
  arXiv:2403.07414}

\bibitem[\protect\citeauthoryear{{M{\'a}rmol-Queralt{\'o}}, {McLure}, {Cullen},
  {Dunlop}, {Fontana}  \& {McLeod}}{{M{\'a}rmol-Queralt{\'o}}
  et~al.}{2016}]{Marmol-Queralto2016}
{M{\'a}rmol-Queralt{\'o}} E.,  {McLure} R.~J.,  {Cullen} F.,  {Dunlop} J.~S.,
  {Fontana} A.,   {McLeod} D.~J.,  2016, \mn@doi [\mnras]
  {10.1093/mnras/stw1212}, \href
  {https://ui.adsabs.harvard.edu/abs/2016MNRAS.460.3587M} {460, 3587}

\bibitem[\protect\citeauthoryear{{Matthee} et~al.,}{{Matthee}
  et~al.}{2023}]{Matthee2023b}
{Matthee} J.,  et~al., 2023, \mn@doi [arXiv e-prints]
  {10.48550/arXiv.2306.05448}, \href
  {https://ui.adsabs.harvard.edu/abs/2023arXiv230605448M} {p. arXiv:2306.05448}

\bibitem[\protect\citeauthoryear{{McLure} et~al.,}{{McLure}
  et~al.}{2018}]{McLure2018}
{McLure} R.~J.,  et~al., 2018, \mn@doi [\mnras] {10.1093/mnras/sty522}, \href
  {https://ui.adsabs.harvard.edu/abs/2018MNRAS.476.3991M} {476, 3991}

\bibitem[\protect\citeauthoryear{{Merlin} et~al.,}{{Merlin}
  et~al.}{2019}]{Merlin2019}
{Merlin} E.,  et~al., 2019, \mn@doi [\mnras] {10.1093/mnras/stz2615}, \href
  {https://ui.adsabs.harvard.edu/abs/2019MNRAS.490.3309M} {490, 3309}

\bibitem[\protect\citeauthoryear{{Nelson} et~al.,}{{Nelson}
  et~al.}{2022}]{Nelson2022}
{Nelson} E.~J.,  et~al., 2022, arXiv e-prints, \href
  {https://ui.adsabs.harvard.edu/abs/2022arXiv220801630N} {p. arXiv:2208.01630}

\bibitem[\protect\citeauthoryear{{Oesch} et~al.,}{{Oesch}
  et~al.}{2023}]{Oesch2023}
{Oesch} P.~A.,  et~al., 2023, \mn@doi [\mnras] {10.1093/mnras/stad2411}, \href
  {https://ui.adsabs.harvard.edu/abs/2023MNRAS.525.2864O} {525, 2864}

\bibitem[\protect\citeauthoryear{{Oke} \& {Gunn}}{{Oke} \&
  {Gunn}}{1983}]{Oke83}
{Oke} J.~B.,  {Gunn} J.~E.,  1983, \mn@doi [\apj] {10.1086/160817}, \href
  {http://adsabs.harvard.edu/abs/1983ApJ...266..713O} {266, 713}

\bibitem[\protect\citeauthoryear{Oliphant}{Oliphant}{2015}]{numpy}
Oliphant T.~E.,  2015, Guide to NumPy.
Continuum Press

\bibitem[\protect\citeauthoryear{{P{\'e}rez-Gonz{\'a}lez}
  et~al.,}{{P{\'e}rez-Gonz{\'a}lez} et~al.}{2023}]{PerezGonzalez2023}
{P{\'e}rez-Gonz{\'a}lez} P.~G.,  et~al., 2023, \mn@doi [\apjl]
  {10.3847/2041-8213/acb3a5}, \href
  {https://ui.adsabs.harvard.edu/abs/2023ApJ...946L..16P} {946, L16}

\bibitem[\protect\citeauthoryear{{P{\'e}rez-Gonz{\'a}lez}
  et~al.,}{{P{\'e}rez-Gonz{\'a}lez} et~al.}{2024}]{PerezGonzalez2024}
{P{\'e}rez-Gonz{\'a}lez} P.~G.,  et~al., 2024, \mn@doi [arXiv e-prints]
  {10.48550/arXiv.2401.08782}, \href
  {https://ui.adsabs.harvard.edu/abs/2024arXiv240108782P} {p. arXiv:2401.08782}

\bibitem[\protect\citeauthoryear{{Rodighiero}, {Bisigello}, {Iani}, {Marasco},
  {Grazian}, {Sinigaglia}, {Cassata}  \& {Gruppioni}}{{Rodighiero}
  et~al.}{2023}]{Rodighiero2023}
{Rodighiero} G.,  {Bisigello} L.,  {Iani} E.,  {Marasco} A.,  {Grazian} A.,
  {Sinigaglia} F.,  {Cassata} P.,   {Gruppioni} C.,  2023, \mn@doi [\mnras]
  {10.1093/mnrasl/slac115}, \href
  {https://ui.adsabs.harvard.edu/abs/2023MNRAS.518L..19R} {518, L19}

\bibitem[\protect\citeauthoryear{{Sanders} et~al.,}{{Sanders}
  et~al.}{2015}]{Sanders2015}
{Sanders} R.~L.,  et~al., 2015, \mn@doi [\apj] {10.1088/0004-637X/799/2/138},
  \href {https://ui.adsabs.harvard.edu/abs/2015ApJ...799..138S} {799, 138}

\bibitem[\protect\citeauthoryear{{Santini} et~al.,}{{Santini}
  et~al.}{2020}]{Santini2021}
{Santini} P.,  et~al., 2020, arXiv e-prints, \href
  {https://ui.adsabs.harvard.edu/abs/2020arXiv201110584S} {p. arXiv:2011.10584}

\bibitem[\protect\citeauthoryear{{Setton} et~al.,}{{Setton}
  et~al.}{2024}]{Setton2024}
{Setton} D.~J.,  et~al., 2024, \mn@doi [arXiv e-prints]
  {10.48550/arXiv.2402.05664}, \href
  {https://ui.adsabs.harvard.edu/abs/2024arXiv240205664S} {p. arXiv:2402.05664}

\bibitem[\protect\citeauthoryear{{Skelton} et~al.,}{{Skelton}
  et~al.}{2014}]{Skelton2014}
{Skelton} R.~E.,  et~al., 2014, \mn@doi [\apjs] {10.1088/0067-0049/214/2/24},
  \href {https://ui.adsabs.harvard.edu/abs/2014ApJS..214...24S} {214, 24}

\bibitem[\protect\citeauthoryear{{Stefanon} et~al.,}{{Stefanon}
  et~al.}{2021}]{Stefanon2021}
{Stefanon} M.,  et~al., 2021, arXiv e-prints, \href
  {https://ui.adsabs.harvard.edu/abs/2021arXiv211006226S} {p. arXiv:2110.06226}

\bibitem[\protect\citeauthoryear{{Valentino} et~al.,}{{Valentino}
  et~al.}{2020}]{Valentino2020}
{Valentino} F.,  et~al., 2020, \mn@doi [\apj] {10.3847/1538-4357/ab64dc}, \href
  {https://ui.adsabs.harvard.edu/abs/2020ApJ...889...93V} {889, 93}

\bibitem[\protect\citeauthoryear{{Valentino} et~al.,}{{Valentino}
  et~al.}{2023}]{Valentino2023}
{Valentino} F.,  et~al., 2023, \mn@doi [\apj] {10.3847/1538-4357/acbefa}, \href
  {https://ui.adsabs.harvard.edu/abs/2023ApJ...947...20V} {947, 20}

\bibitem[\protect\citeauthoryear{{Virtanen} et~al.,}{{Virtanen}
  et~al.}{2020}]{scipy}
{Virtanen} P.,  et~al., 2020, \mn@doi [Nature Methods]
  {10.1038/s41592-019-0686-2}, \href
  {https://ui.adsabs.harvard.edu/abs/2020NatMe..17..261V} {17, 261}

\bibitem[\protect\citeauthoryear{{Wang} et~al.,}{{Wang}
  et~al.}{2016}]{Wang2016}
{Wang} T.,  et~al., 2016, \mn@doi [\apj] {10.3847/0004-637X/816/2/84}, \href
  {https://ui.adsabs.harvard.edu/abs/2016ApJ...816...84W} {816, 84}

\bibitem[\protect\citeauthoryear{{Wang} et~al.,}{{Wang}
  et~al.}{2019}]{Wang2019}
{Wang} T.,  et~al., 2019, \mn@doi [\nat] {10.1038/s41586-019-1452-4}, \href
  {https://ui.adsabs.harvard.edu/abs/2019Natur.572..211W} {572, 211}

\bibitem[\protect\citeauthoryear{{Wang} et~al.,}{{Wang}
  et~al.}{2024}]{Wang2024}
{Wang} T.,  et~al., 2024, \mn@doi [arXiv e-prints] {10.48550/arXiv.2403.02399},
  \href {https://ui.adsabs.harvard.edu/abs/2024arXiv240302399W} {p.
  arXiv:2403.02399}

\bibitem[\protect\citeauthoryear{{Weibel} et~al.,}{{Weibel}
  et~al.}{2024}]{Weibel2024}
{Weibel} A.,  et~al., 2024, \mn@doi [arXiv e-prints]
  {10.48550/arXiv.2403.08872}, \href
  {https://ui.adsabs.harvard.edu/abs/2024arXiv240308872W} {p. arXiv:2403.08872}

\bibitem[\protect\citeauthoryear{{Whitaker}, {Pope}, {Cybulski}, {Casey},
  {Popping}  \& {Yun}}{{Whitaker} et~al.}{2017}]{Whitaker2017}
{Whitaker} K.~E.,  {Pope} A.,  {Cybulski} R.,  {Casey} C.~M.,  {Popping} G.,
  {Yun} M.~S.,  2017, \mn@doi [\apj] {10.3847/1538-4357/aa94ce}, \href
  {https://ui.adsabs.harvard.edu/abs/2017ApJ...850..208W} {850, 208}

\bibitem[\protect\citeauthoryear{{Williams} et~al.,}{{Williams}
  et~al.}{2014}]{Williams2014}
{Williams} C.~C.,  et~al., 2014, \mn@doi [\apj] {10.1088/0004-637X/780/1/1},
  \href {https://ui.adsabs.harvard.edu/abs/2014ApJ...780....1W} {780, 1}

\bibitem[\protect\citeauthoryear{{Williams} et~al.,}{{Williams}
  et~al.}{2019}]{Williams2019}
{Williams} C.~C.,  et~al., 2019, \mn@doi [\apj] {10.3847/1538-4357/ab44aa},
  \href {https://ui.adsabs.harvard.edu/abs/2019ApJ...884..154W} {884, 154}

\bibitem[\protect\citeauthoryear{{Williams} et~al.,}{{Williams}
  et~al.}{2023}]{Williams2023}
{Williams} C.~C.,  et~al., 2023, \mn@doi [arXiv e-prints]
  {10.48550/arXiv.2311.07483}, \href
  {https://ui.adsabs.harvard.edu/abs/2023arXiv231107483W} {p. arXiv:2311.07483}

\bibitem[\protect\citeauthoryear{{Xiao} et~al.,}{{Xiao}
  et~al.}{2023a}]{Xiao2023}
{Xiao} M.,  et~al., 2023a, \mn@doi [arXiv e-prints]
  {10.48550/arXiv.2309.02492}, \href
  {https://ui.adsabs.harvard.edu/abs/2023arXiv230902492X} {p. arXiv:2309.02492}

\bibitem[\protect\citeauthoryear{{Xiao} et~al.,}{{Xiao}
  et~al.}{2023b}]{Xiao2023a}
{Xiao} M.~Y.,  et~al., 2023b, \mn@doi [\aap] {10.1051/0004-6361/202245100},
  \href {https://ui.adsabs.harvard.edu/abs/2023A&A...672A..18X} {672, A18}

\bibitem[\protect\citeauthoryear{{Yamaguchi} et~al.,}{{Yamaguchi}
  et~al.}{2019}]{Yamaguchi2019}
{Yamaguchi} Y.,  et~al., 2019, \mn@doi [\apj] {10.3847/1538-4357/ab0d22}, \href
  {https://ui.adsabs.harvard.edu/abs/2019ApJ...878...73Y} {878, 73}

\bibitem[\protect\citeauthoryear{{Zavala} et~al.,}{{Zavala}
  et~al.}{2023}]{Zavala2023}
{Zavala} J.~A.,  et~al., 2023, \mn@doi [\apjl] {10.3847/2041-8213/acacfe},
  \href {https://ui.adsabs.harvard.edu/abs/2023ApJ...943L...9Z} {943, L9}

\makeatother
\end{thebibliography}

\appendix
\makeatletter
\renewcommand{\fps@figure}{htbp}
\renewcommand{\fps@table}{htbp}
\makeatother

\section{2D spectra of our sample}
\label{Appendix}

Here, we present the spectra of the 19 HST-dark galaxies in our sample, which constitute the main targets of the cycle 1 JWST/NIRSpec program: `Quiescent or dusty? Unveiling the nature of red galaxies at z > 3'
(GO-2198; PIs: L. Barrufet \& P. Oesch).

\begin{figure}
 \centering
\includegraphics[width=\columnwidth]{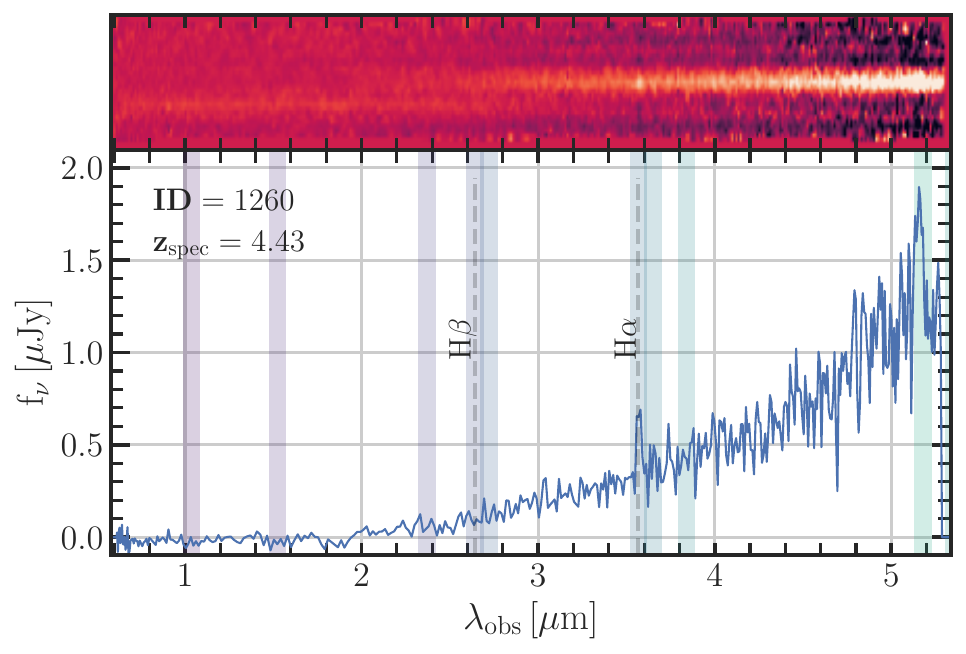}   
\includegraphics[width=\columnwidth]{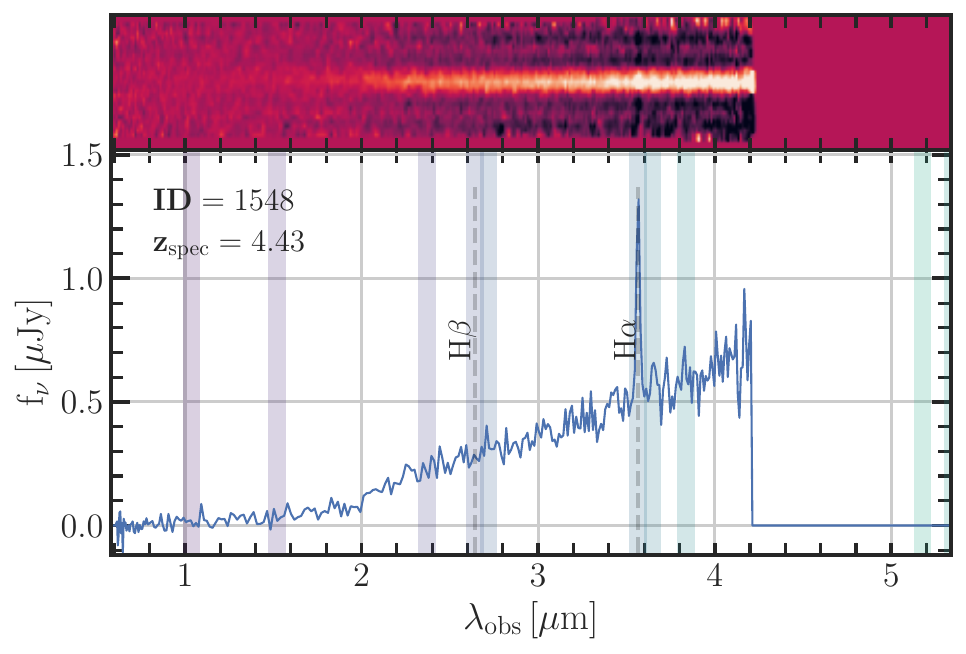}   
\caption{2D and 1D Spectra for the HST-dark galaxies. The dashed lines mark the locations of the $\mathrm{H\alpha}$ and $\mathrm{H\beta}$ spectral lines. Coloured lines represent [OIII], [SII], HeI, and [CI], respectively. The ID and the spectroscopic redshift are shown in the upper left of the  1D plot} 
   \label{Spectra1}
\end{figure}

\begin{figure}
 \centering
 \includegraphics[width=\columnwidth]{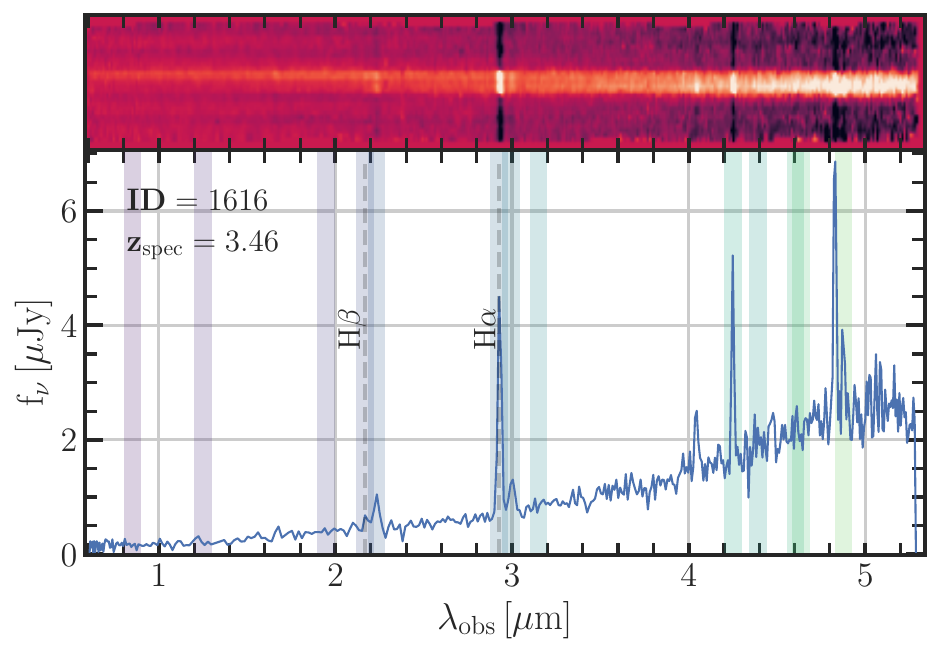}
 \includegraphics[width=\columnwidth]{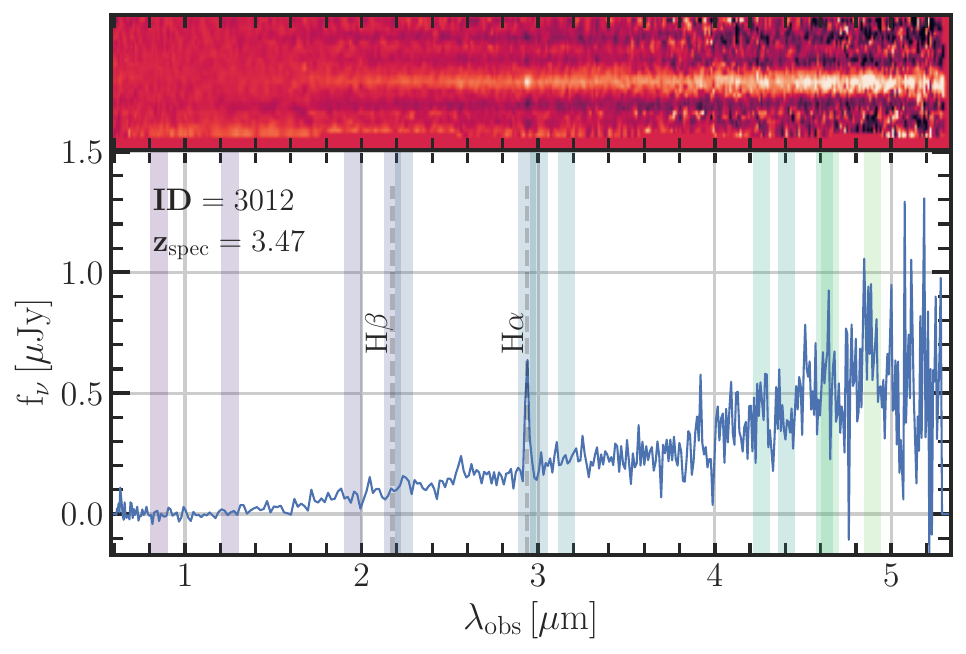}
\includegraphics[width=\columnwidth]{Figures/2DSpectra3339.pdf}   
\caption{2D and 1D Spectra for the HST-dark galaxies. The dashed lines mark the locations of the $\mathrm{H\alpha}$ and $\mathrm{H\beta}$ spectral lines. Coloured lines represent [OIII], [SII], HeI, and [CI], respectively. The ID and the spectroscopic redshift are shown in the upper left of the  1D plot} 
   \label{Spectra2}
\end{figure}

\begin{figure}
 \centering
\includegraphics[width=\columnwidth]{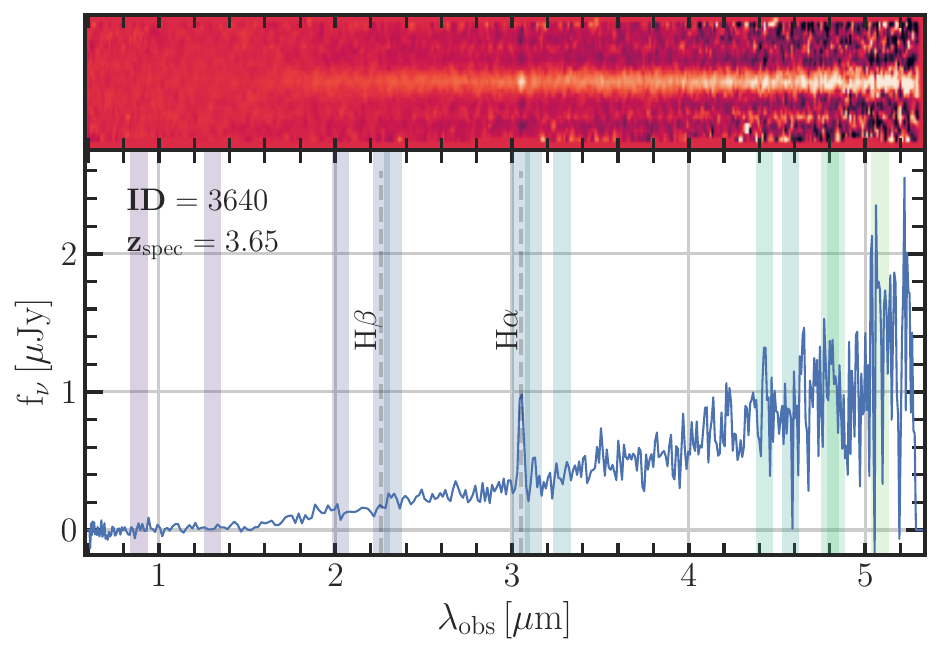}   
\includegraphics[width=\columnwidth]{Figures/2DSpectra4307.pdf}  
\includegraphics[width=\columnwidth]{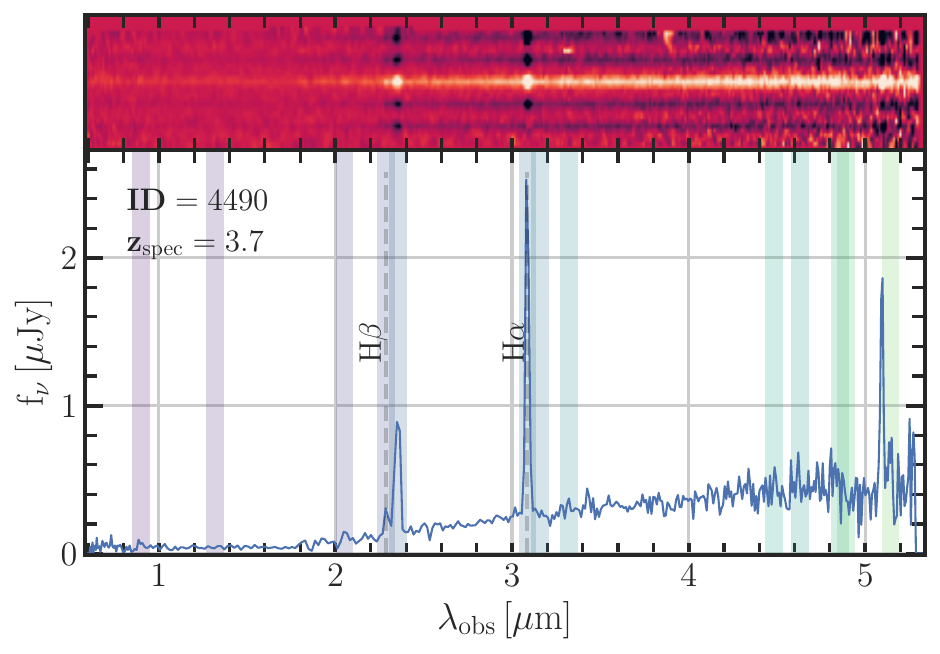}
\caption{2D and 1D Spectra for the HST-dark galaxies. The dashed lines mark the locations of the $\mathrm{H\alpha}$ and $\mathrm{H\beta}$ spectral lines. Coloured lines represent [OIII], [SII], HeI, and [CI], respectively. The ID and the spectroscopic redshift are shown in the upper left of the  1D plot} 
   \label{Spectra3}
\end{figure}

\begin{figure}
 \centering
\includegraphics[width=\columnwidth]{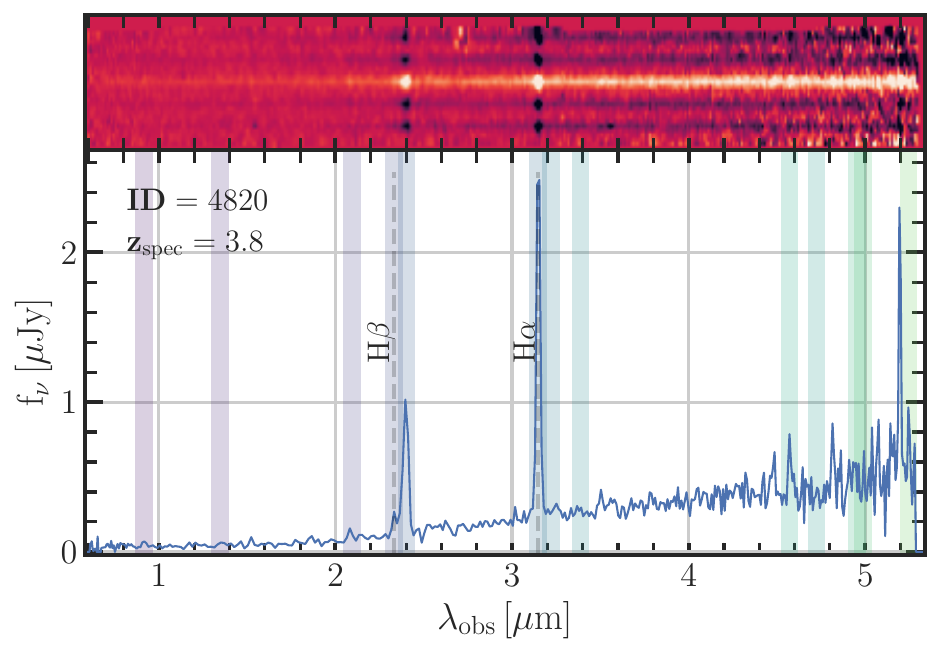}   
\includegraphics[width=\columnwidth]{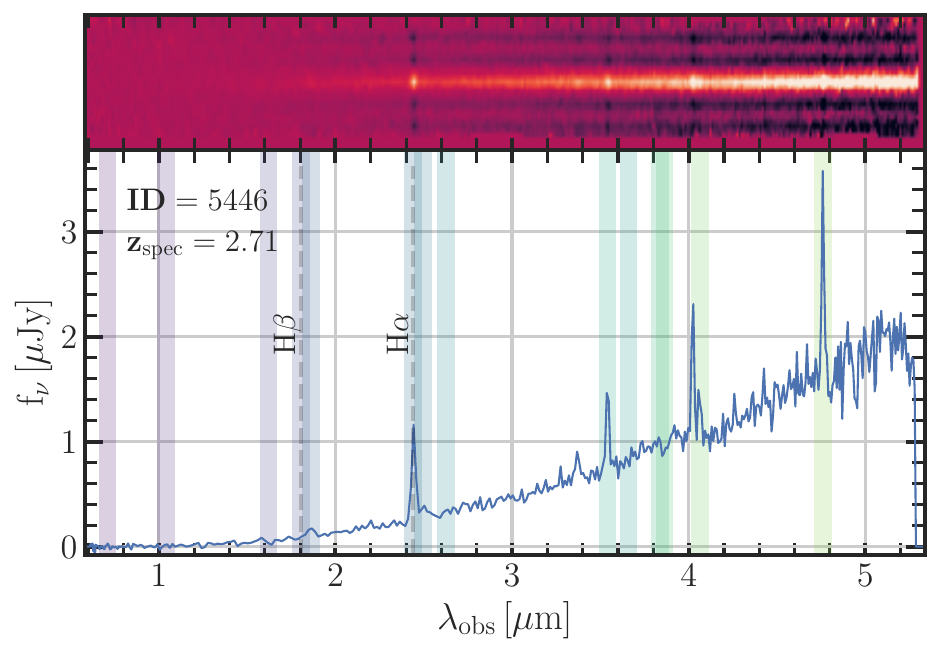}   
\includegraphics[width=\columnwidth]{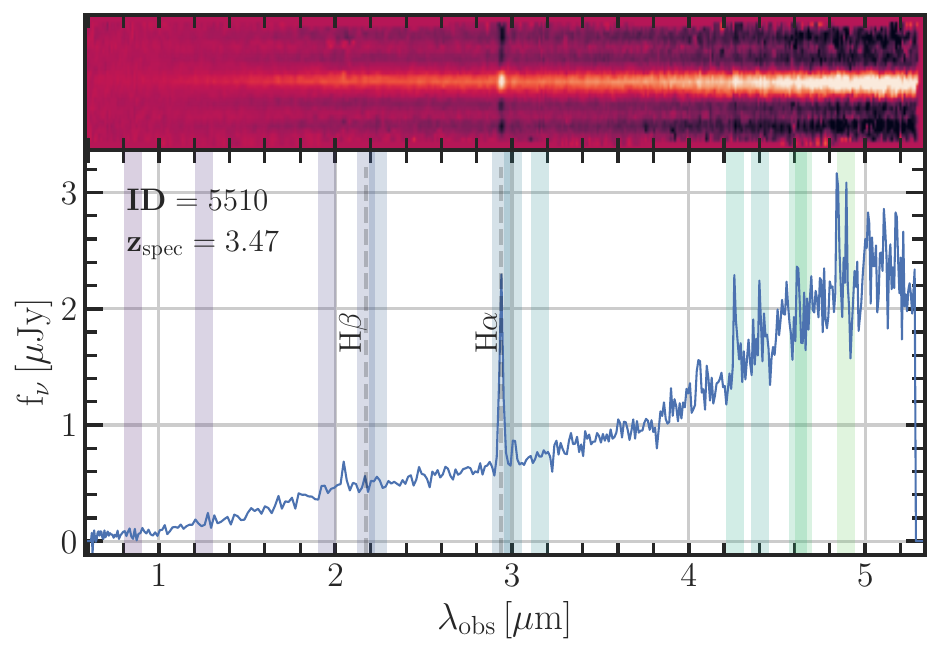} 
\caption{2D and 1D Spectra for the HST-dark galaxies} 
   \label{Spectra4}
\end{figure}

\begin{figure}
 \centering
\includegraphics[width=\columnwidth]{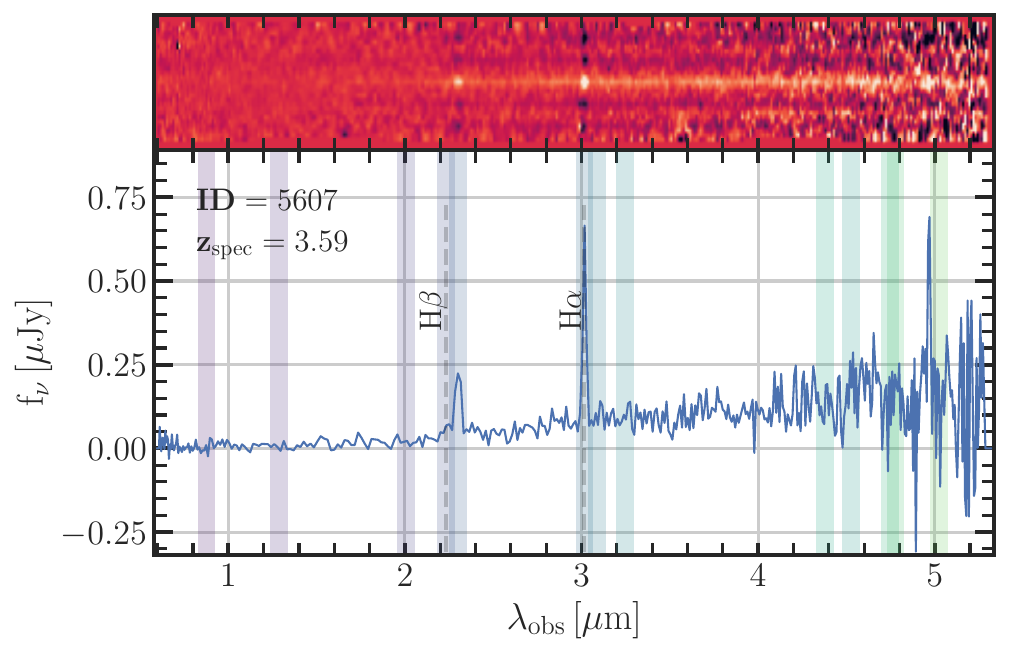}
\includegraphics[width=\columnwidth]{Figures/2DSpectra6151.pdf}   
\includegraphics[width=\columnwidth]{Figures/2DSpectra6620.pdf}   
\caption{2D and 1D Spectra for the HST-dark galaxies. The dashed lines mark the locations of the $\mathrm{H\alpha}$ and $\mathrm{H\beta}$ spectral lines. Coloured lines represent [OIII], [SII], HeI, and [CI], respectively. The ID and the spectroscopic redshift are shown in the upper left of the  1D plot.} 
   \label{Spectra5}
\end{figure}

\begin{figure}
 \centering
\includegraphics[width=\columnwidth]{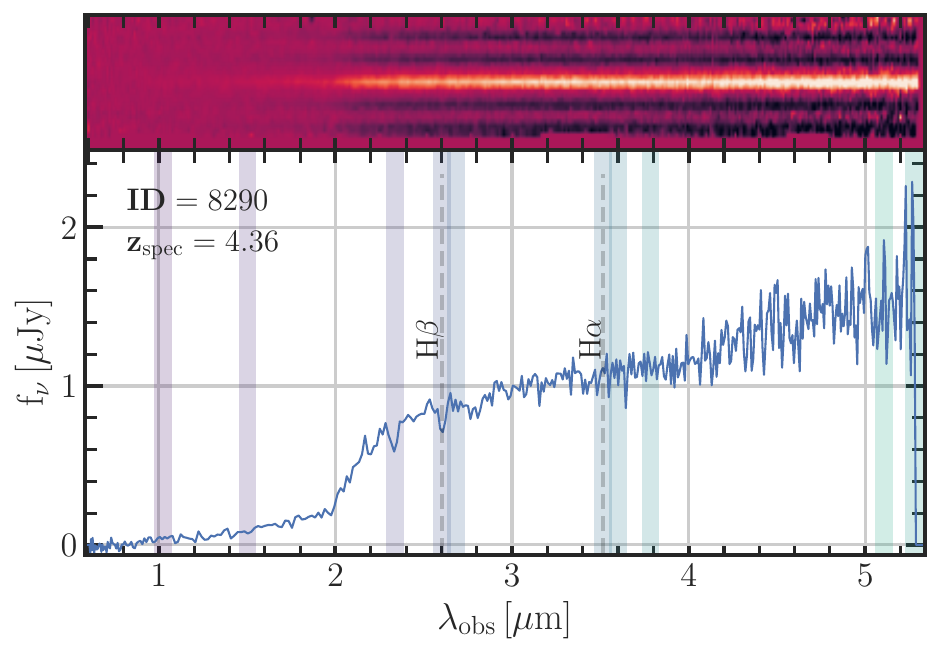}
\includegraphics[width=\columnwidth]{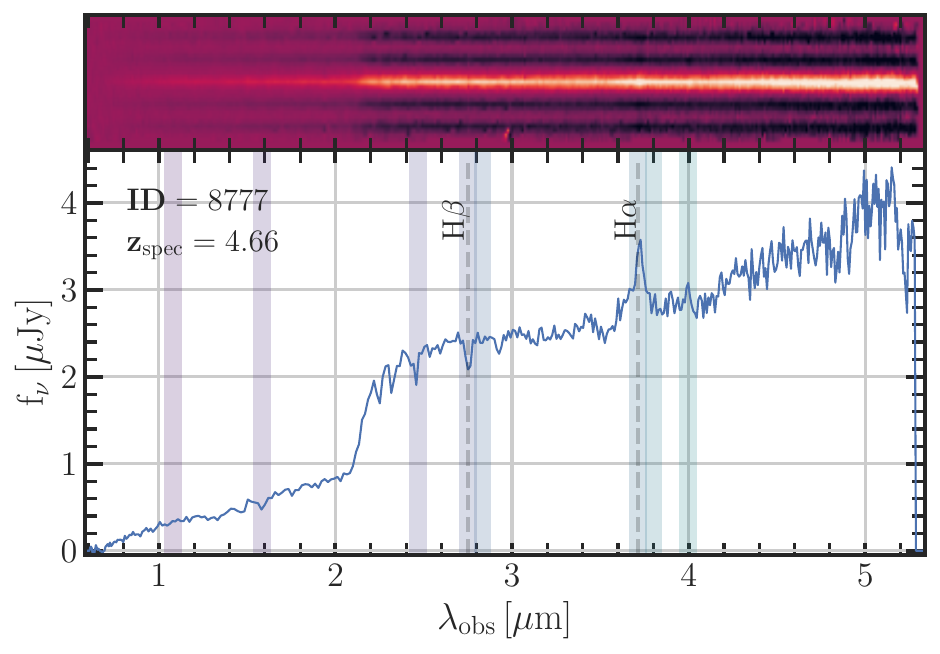}   
\caption{2D and 1D Spectra for the HST-dark galaxies. The dashed lines mark the locations of the $\mathrm{H\alpha}$ and $\mathrm{H\beta}$ spectral lines. Coloured lines represent [OIII], [SII], HeI, and [CI], respectively. The ID and the spectroscopic redshift are shown in the upper left of the  1D plot.}  
   \label{Spectra6}
\end{figure}

\begin{figure}
 \centering
\includegraphics[width=\columnwidth]{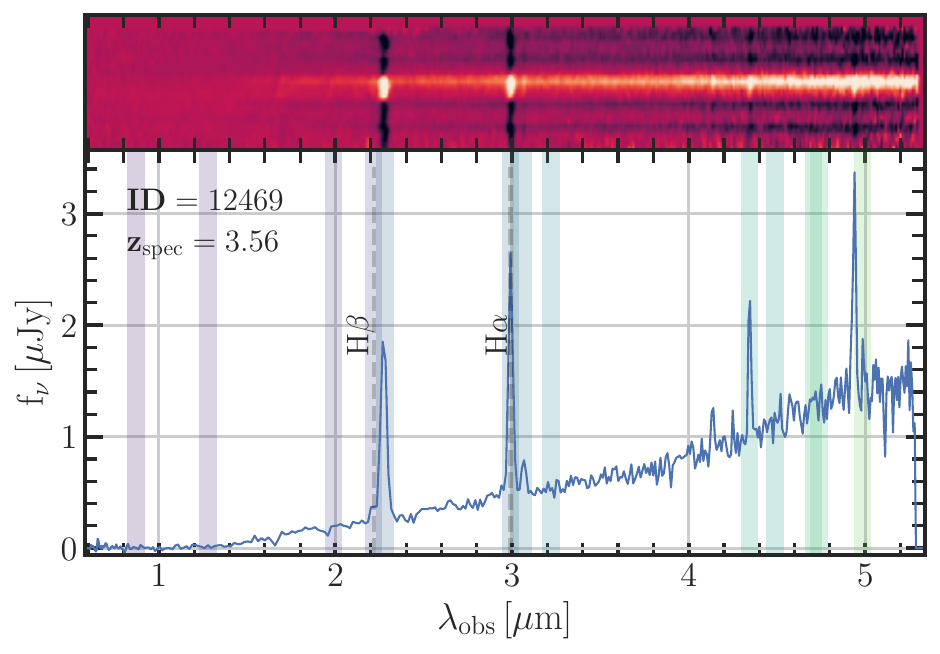}   
\includegraphics[width=\columnwidth]{Figures/2DSpectra12577.pdf} 
\caption{2D and 1D Spectra for the HST-dark galaxies. The dashed lines mark the locations of the $\mathrm{H\alpha}$ and $\mathrm{H\beta}$ spectral lines. Coloured lines represent [OIII], [SII], HeI, and [CI], respectively. The ID and the spectroscopic redshift are shown in the upper left of the  1D plot.} 
   \label{Spectra7}
\end{figure}

\bsp	
\label{lastpage}
\end{document}